\documentclass[11pt]{article}

\usepackage{a4wide}

\usepackage{amsmath}
\usepackage{amssymb}
\usepackage{amsfonts}
\usepackage{amsthm}
\usepackage{array}
\usepackage{float}
\usepackage{xy}
\usepackage{enumerate}
\usepackage{graphicx}

\newcolumntype{C}[1]{>{\centering\arraybackslash$}m{#1}<{$}}
\newlength{\mycolwd}

\newcommand{\rmi}{\mathrm{i}}

\begin{document}

\begin{center}

{\LARGE{Nongeneric positive partial transpose states\\
of rank five in $3\times 3$ dimensions}}

\bigskip

{\large{Leif Ove Hansen and Jan Myrheim\\
Department of Physics, Norwegian University of Science and Technology,\\
N-7491 Trondheim, Norway}}

\bigskip

{\large{\today}}

\end{center}

\bigskip
\bigskip

\begin{abstract}

  In $3\times 3$ dimensions, entangled mixed states that are positive
  under partial transposition (PPT states) must have rank at least
  four. These rank four states are completely understood.  We say that
  they have rank $(4,4)$ since a state $\rho$ and its partial
  transpose $\rho^P$ both have rank four.  The next problem is to
  understand the extremal PPT states of rank $(5,5)$.  We call two
  states $\textrm{SL}\otimes\textrm{SL}$-equivalent if they are
  related by a product transformation.  A generic rank $(5,5)$ PPT
  state $\rho$ is extremal, and $\rho$ and $\rho^P$ both have six
  product vectors in their ranges, and no product vectors in their
  kernels.  The three numbers $\{6,6;0\}$ are
  $\textrm{SL}\otimes\textrm{SL}$-invariants that help us classify the
  state.  There is no analytical understanding of such states.  We
  have studied numerically a few types of nongeneric rank five PPT
  states, in particular states with one or more product vectors in
  their kernels.  We find an interesting new analytical construction
  of all rank four extremal PPT states, up to
  $\textrm{SL}\otimes\textrm{SL}$-equivalence, where they appear as
  boundary states on one single five dimensional face on the set of
  normalized PPT states.  The interior of the face consists of rank
  $(5,5)$ states with four common product vectors in their kernels, it
  is a simplex of separable states surrounded by entangled PPT states.
  We say that a state $\rho$ is
  $\textrm{SL}\otimes\textrm{SL}$-symmetric if $\rho$ and $\rho^P$ are
  $\textrm{SL}\otimes\textrm{SL}$-equivalent, and is genuinely
  $\textrm{SL}\otimes\textrm{SL}$-symmetric if it is
  $\textrm{SL}\otimes\textrm{SL}$-equivalent to a state $\tau$ with
  $\tau=\tau^P$.  Genuine $\textrm{SL}\otimes\textrm{SL}$-symmetry
  implies a special form of $\textrm{SL}\otimes\textrm{SL}$-symmetry.
  We have produced numerically, by a special method, a random sample
  of rank $(5,5)$ $\textrm{SL}\otimes\textrm{SL}$-symmetric states.
  About fifty of these are of type $\{6,6;0\}$, among those all are
  extremal and about half are genuinely
  $\textrm{SL}\otimes\textrm{SL}$-symmetric.  All these genuinely
  $\textrm{SL}\otimes\textrm{SL}$-symmetric states can be transformed
  to have a circulant form.  We find however that this is not a
  generic property of genuinely
  $\textrm{SL}\otimes\textrm{SL}$-symmetric states.  The remaining
  $\textrm{SL}\otimes\textrm{SL}$-symmetric states found in the search
  have product vectors in their kernels, they inspired us to study
  such states without regard to
  $\textrm{SL}\otimes\textrm{SL}$-symmetry.

\end{abstract}


\section{Introduction}

Entanglement between subsystems of a composite quantum system is a
phenomenon which has no counterpart in classical physics. Entangled
quantum states show correlations in measurements which cannot be
modelled within any local theory, including classical physics. A
classical local description of such systems implies so called Bell
inequalities \cite{Bell64}, which are violated in
experiments~\cite{Aspect82}. So entangled quantum states exhibit a
nonlocality that excludes any local theory, though nonlocal
deterministic theories are still possible.

Pure product states are the only pure quantum states that are not
entangled, and they resemble pure classical states in that they have
no correlations at all. By definition, a mixed quantum state is a
statistical ensemble of pure quantum states, and it is represented
mathematically by a density matrix.  One single density matrix may
represent many different ensembles.  A basic postulate is that there
is no way to distinguish experimentally between different ensembles
represented by the same density matrix.

A mixed quantum state is said to be separable if it can be mixed
entirely from pure product states. The entangled states are exactly
those that are not separable.  While the separability problem for pure
states is solved entirely via Schmidt decomposition of state vectors,
the problem of how to characterize the set $\mathcal{S}$ of separable
mixed states, and to decide whether a given mixed state is separable
or entangled, is known to be a very difficult mathematical problem in
general, and it has been demonstrated that operational procedures are
NP-hard~\cite{Gharibian10}.

In recent years these problems have been given considerable attention,
mainly due to the fact that quantum entanglement has found use in many
applications. Many new developments require an understanding of
entanglement as a resource, in areas such as quantum communication,
quantum cryptography, and quantum computing.  How to prepare,
manipulate and detect entangled quantum states has become an important
issue.

Sound operational methods or criteria for entanglement checking only
exist for special cases and/or for low dimensional systems. The
separable states have the property that they remain positive after
partial transposition, they are PPT states. The set $\mathcal{P}$ of
PPT states is in general larger than the set $\mathcal{S}$ of
separable states, but the difference between the two sets is small in
low dimensions, and in the $2\times 2$ and $2\times 3$ systems
$\mathcal{P}=\mathcal{S}$ \cite{MPRHorodecki96}. Thus the condition of
positive partial transpose, known as the Peres separability criterion
\cite{Peres96}, is completely adequate as long as the system has
dimension $N=N_AN_B\leq 6$.

For systems of dimension $N=N_AN_B\geq 8$, entangled PPT states, also
known as states with bound entanglement, exist. It is these states
that make up the difference between $\mathcal{P}$ and $\mathcal{S}$,
they are the states where the Peres separability criterion is not
sufficient.

Another useful criterion is the range criterion, which states that if
a state is separable then its range is spanned by a set of product
vectors $w_i=u_i\otimes v_i$ such that the range of the partial
transpose $\rho^P$ is spanned by $\widetilde{w}_i=u_i\otimes v_i^*$.
It was shown that this criterion is independent from the PPT
criterion, as there are PPT entangled states that violate the range
criterion and there are NPT states satisfying it~\cite{PHorodecki97}.

The close relation between PPT states and product vectors has been
used to prove the separability of sufficiently low rank PPT states.
It was shown that all PPT states of rank at most $N_B$ supported on a
Hilbert space of dimension $N=N_AN_B$ with $N_A\leq N_B$, are
separable~\cite{PHorodecki00}.

For the $3\times 3$ system this means that the lowest possible rank of
an entangled PPT state is four. Bennett \emph{et
  al.}~\cite{Bennett99,DiVincenzo03} introduced a method for
constructing low rank entangled PPT states by using unextendable
product bases (UPBs). A UPB is defined as a maximal set of orthogonal
product vectors which is not a complete basis of the Hilbert space. If
one constructs an orthogonal projection $Q$ onto the UPB, then its
complementary projection $P=I-Q$ is an entangled PPT state.  The UPB
construction is most successful in the case of rank four PPT states in
the $3\times 3$ system, where it leads to a construction of all
entangled PPT states of rank four \cite{Leinaas10A,Chen11}.

This UPB strategy fails for rank five states in the $3\times 3$
system, since a UPB must lie in the kernel of the state it defines,
and no UPB has only four elements.  The lack of such a simple
construction method makes the characterization of entangled PPT states
in the $3\times 3$ system a much more challenging problem for the rank
five states than for the rank four states.  We report here on a study
of the rank five states, describing in particular several nongeneric
forms.

A PPT state is partly characterized by two numbers $(m,n)$, where $m$
is its rank and $n$ is the rank of its partial transpose.  A necessary
condition for a PPT state in dimension $3\times 3$ to be extremal is
that $m^2+n^2\leq 82$~\cite{Leinaas07}.  Most of the states we have
studied have rank $(5,5)$, but it should be noted that states of rank
$(5,6)$ and rank $(5,7)$ are also easy to find numerically, and they are
generically extremal.  There also exist PPT states of rank $(5,8)$,
but they can not be extremal~\cite{Leinaas10B}.

\subsubsection*{Outline of the paper}

The contents of the present paper are organized in the following
manner.

In Sections~\ref{BasicLinearAlgebra} and~\ref{sec:compositesystems} we
review some linear algebra and introduce notation.  We emphasize the
importance of product vectors and product transformations in our study
of low-rank PPT states in bipartite composite systems.  It is useful
to classify such a state by the number of product vectors in its range
and kernel.

Since a generic subspace of dimension four contains no product
vectors, a rank five state with product vectors in its kernel is
nongeneric.  In Section~\ref{sec:rank44revisited} we define a standard
form for PPT states of rank five with four product vectors in the
kernel.  All these special PPT states of rank $(5,5)$ are nonextremal,
but this scheme leads to a new method for constructing extremal PPT
states of rank $(4,4)$, and is highly relevant to our further study of
nongeneric PPT states of rank $(5,5)$, presented in
Section~\ref{Nongeneric55}.

In Section~\ref{sec:generic55} we review some features regarding
product vectors in generic subspaces of dimension five in the
$3\times 3$ system.  We also discuss nongeneric subspaces in the
$3\times 3$ system, specifically how to construct pairs of orthogonal
subspaces $\mathcal{U}$ and $\mathcal{V}$ with $|\mathcal{U}|=5$ and
$|\mathcal{V}|=4$, such that the number of product vectors in
$\mathcal{V}$ is nonzero.

In Section~\ref{NumericalResults} a summary of our numerical results
on rank five PPT states is presented. This includes data for the
generic states $\rho$ with no product vectors in $\mathrm{Ker}\,\rho$,
and several nongeneric cases with up to four product vectors in
$\mathrm{Ker}\,\rho$. We also present some results from our random
searches for $\textrm{SL}\otimes\textrm{SL}$-symmetric states.

In Section~\ref{Nongeneric55} we present a collection of nongeneric
standard forms for orthogonal subspaces $\mathcal{U}$ and
$\mathcal{V}$, with dimensions five and four respectively. The
nongeneric feature is that the number of product vectors in
$\mathcal{V}$ is nonzero. For the various $\mathcal{U}$ and
$\mathcal{V}$, we have produced PPT states of rank $(5,5)$ with
$\mathrm{Img}\,\rho=\mathcal{U}$ and
$\mathrm{Ker}\,\rho=\mathcal{V}$. The number of product vectors in
$\mathrm{Img}\,\rho$ is either six or infinite, while the number of
product vectors in $\mathrm{Ker}\,\rho$ ranges from one to four. The
latter is presented as a special case in
Section~\ref{sec:rank44revisited}.

Finally, in Section~\ref{sec:Averyspecialsubspace} we present a very
special five-dimensional subspace of ${\mathbb{C}^9}$ that contains
only two product vectors, with one product vector in the orthogonal
complement.  We construct analytically a set of states found by
transformation to a standard form of one particular rank $(5,5)$
nonextremal PPT state in this subspace.  This nonextremal PPT state
was found completely by chance in our random searches for
$\textrm{SL}\otimes\textrm{SL}$-symmetric states.


\section{Basic linear algebra}
\label{BasicLinearAlgebra}

We want to review some basic concepts of linear algebra, partly in
order to define our notation and make the paper self-contained, and
partly in order to review some useful but less well known facts.

\subsection{Density matrices}

The natural structure of the set $H_N$ of Hermitian $N\times N$
matrices is that of a real Hilbert space of dimension $N^2$ with the
scalar product
\begin{equation}
	(X,Y)=\textrm{Tr}\,(XY)\;.
\end{equation}
The set of mixed states, or density matrices, is defined as
\begin{equation}
  \mathcal{D}=\mathcal{D}_N
  =\{\rho\in H_N\,|\,\rho \geq 0,\,\textrm{Tr}\,\rho=1\,\}\;.
\end{equation}
A density matrix $\rho$ has a spectral representation in terms of a
complete set of orthonormal eigenvectors $\psi_i\in {\mathbb{C}^N}$
with eigenvalues $\lambda_i\geq 0$,
\begin{equation}
	\rho=\sum_{i=1}^{N}\,\lambda_i\,\psi_i{\psi_i}^{\dagger}\;,
	\qquad\qquad
        {\psi_i}^{\dagger}\psi_j=\delta_{ij}\;,
	\qquad\qquad
	\sum_{i=1}^{N}\,\psi_i{\psi_i}^{\dagger} = {I}\;.
\end{equation}
The spectral representation is one particular ensemble representation
of $\rho$.  It implies that
\begin{equation}
  \mathrm{Tr}\,\rho=\sum_{i=1}^{N}\,\lambda_i\;.
\end{equation}
The rank of $\rho$ is the number of eigenvalues $\lambda_i>0$.  The
matrices
\begin{equation}
	\label{PQ}
	P=\sum_{i,\,\lambda_i>0}\,\psi_i{\psi_i}^{\dagger} \;,
	\qquad\qquad 
	Q={I}-P = \sum_{i,\,\lambda_i= 0}\,\psi_i{\psi_i}^{\dagger}
\end{equation}
are Hermitian and project orthogonally onto the two complementary
orthogonal subspaces $\textrm{Img}\,\rho$, the range of $\rho$, and
$\textrm{Ker}\,\rho$, the kernel (or null space) of $\rho$.

When all $\lambda_i \geq 0$ we say that $\rho$ is positive (or
positive semidefinite) and write $\rho\geq 0$.  An equivalent
condition is that ${\psi}^{\dagger}\rho{\psi}\geq 0$ for all
$\psi \in {\mathbb{C}^N}$.  It follows from the last inequality and
the spectral representation of $\rho$ that
${\psi}^{\dagger}\rho{\psi}=0\Leftrightarrow \rho\psi=0$.

The fact that the positivity conditions
${\psi}^{\dagger}\rho{\psi}\geq 0$ are linear in $\rho$ implies that
$\mathcal{D}$ is a convex set, so that if $\rho$ is a proper convex
combination of $\rho_1,\rho_2\in{\mathcal{D}}$,
\begin{equation}
	\rho=p\rho_1+(1-p)\rho_2\;,
	\qquad\qquad
	0<p<1\;,
\end{equation}
then $\rho\in{\mathcal{D}}$.  Furthermore, since
\begin{equation}
\textrm{Ker}\,\rho
=\{\psi\,|\,\rho\psi=0\}
=\{\psi\,|\,{\psi}^{\dagger}\rho\psi=0\}
\end{equation}
when $\rho\geq 0$, it follows that
\begin{equation}
	\textrm{Ker}\,\rho=\textrm{Ker}\,\rho_1 \cap \textrm{Ker}\,\rho_2
\end{equation}
independent of $p$, when $\rho$ is a proper convex combination as
above.  Since $\textrm{Ker}\,\rho$ is independent of $p$, so is
$\textrm{Img}\,\rho$ = $(\textrm{Ker}\,\rho)^{\perp}$.

A compact (closed and bounded) convex set is determined by its
extremal points, those points that are not convex combinations of
other points in the set.  The extremal points of $\mathcal{D}$ are the
pure states of the form $\rho=\psi\psi^{\dagger}$ with
$\psi\in {\mathbb{C}^N}$ and $\psi^{\dagger}\psi=1$.  Thus the
spectral representation is an expansion of $\rho$ as a convex
combination of $m$ extremal points in ${\mathcal{D}}$ where $m$ is the
rank of $\rho$.

\subsection{Perturbations and extremality in $\mathcal{D}$}

Let $\rho$ be a density matrix, and define the projections $P$ and $Q$
as in Equation~\eqref{PQ}.  Consider a perturbation of the form
\begin{equation}
\label{eq:pertrho}
  \rho\to\rho'=\rho+\epsilon A
\end{equation}
where $A\neq 0$ is Hermitian, and $\textrm{Tr}\,A=0$ so that
$\textrm{Tr}\,\rho\,'=\textrm{Tr}\,\rho$.  The real parameter
$\epsilon$ may be infinitesimal or finite.

We observe that if $\textrm{Img}\,A \subset \textrm{Img}\,\rho$, or
equivalently if $PAP=A$, then there will be a finite range of values
of $\epsilon$, say $\epsilon_1\leq\epsilon\leq\epsilon_2$ with
$\epsilon_1 < 0 < \epsilon_2$, such that $\rho\,'\in{\mathcal{D}}$ and
$\textrm{Img}\,\rho = \textrm{Img}\,\rho\,'$. This is so because the
eigenvectors of $\rho$ with zero eigenvalue will remain eigenvectors
of $\rho\,'$ with zero eigenvalue, and all the positive eigenvalues of
$\rho$ will change continuously with $\epsilon$ into eigenvalues of
$\rho\,'$.  Since $\mathcal{D}$ is compact, we may choose $\epsilon_1$
and $\epsilon_2$ such that $\rho\,'$ has at least one negative
eigenvalue when either $\epsilon<\epsilon_1$ or $\epsilon>\epsilon_2$.
The negative eigenvalue becomes zero at $\epsilon=\epsilon_1$ or
$\epsilon=\epsilon_2$, this makes $\textrm{Ker}\,\rho\,'$ strictly
larger than $\textrm{Ker}\,\rho$ and $\textrm{Img}\,\rho\,'$ strictly
smaller than $\textrm{Img}\,\rho$ in both limits $\epsilon=\epsilon_1$
and $\epsilon=\epsilon_2$.

The other way around, if $\rho\,'\in\mathcal{D}$ for
$\epsilon_1\leq\epsilon\leq\epsilon_2$ with $\epsilon_1 < 0 <
\epsilon_2$, then $\rho\,'$ is a convex combination of
$\rho+\epsilon_1 A$ and $\rho+\epsilon_2 A$ for every $\epsilon$ in
the open interval $\epsilon_1 < \epsilon < \epsilon_2$. Hence
$\textrm{Img}\,\rho\,'$ is independent of $\epsilon$ in this open
interval, implying that $\textrm{Img}\,A \subseteq \textrm{Img}\,\rho$
and $PAP=A$.

This gives us three equivalent formulations for the extremality
condition of $\rho$ on $\mathcal{D}$. The state $\rho$ is extremal in
$\mathcal{D}$ if and only if

\begin{itemize}

\item[--]
  There exists no $A\neq 0$ with $\textrm{Tr}\,A=0$ and $PAP=A$.

\item[--]
  The equation $PAP=A$ for the Hermitian matrix $A$ has $A=\rho$ as
  its only solution (up to proportionality).

\item[--]
  There exists no $\rho\,'\in \mathcal{D}$ with $\rho\,'\neq \rho$ and
  $\textrm{Img}\,\rho\,'=\textrm{Img}\,\rho$.

\end{itemize}

We may replace the condition $PAP=A$ by the weaker condition $QAQ=0$.
By standard perturbation theory it implies that the zero eigenvalues
of $\rho$ do not change to first order in $\epsilon$.  Thus, the
perturbation in Equation~\eqref{eq:pertrho}, with $QAQ=0$ and
$\epsilon$ infinitesimal, preserves the rank but not necessarily the
range of $\rho$.

\subsection{Projection operators on $H_N$}

Using the projections $P$ and $Q$ defined above we can define
projection operators ${\mathbf{P}}$, ${\mathbf{Q}}$, and ${\mathbf{R}}$ on
$H_N$, the real Hilbert space of Hermitian $N\times N$ matrices, as
follows,
\begin{eqnarray}
	{\mathbf{P}}X & \!\!\! = & \!\!\! PXP\;,\nonumber\\ 
	{\mathbf{Q}}X & \!\!\! = & \!\!\! QXQ\;,\nonumber\\
	{\mathbf{R}}X & \!\!\! = & \!\!\! ({\mathbf{I-P-Q}})X\;.
\end{eqnarray}
Here ${\mathbf{I}}$ is the identity operator on $H_N$. It is
straightforward to verify that these are complementary projections,
with
${\mathbf{P^{\textnormal{2}}=P,\,Q^{\textnormal{2}}=Q,\,PQ=QP=0}}$,
and so on. They are symmetric with respect to the natural scalar
product on $H_N$, hence they project orthogonally, and relative to an
orthonormal basis for $H_N$ they are represented by symmetric
matrices.

Relative to an orthonormal basis of ${\mathbb{C}^N}$ with the first
$m$ basis vectors in $\textrm{Img}\,\rho$ and the last $k=N-m$ basis
vectors in $\textrm{Ker}\,\rho$, a Hermitian matrix $X$ takes the
block form
\begin{equation}
	X=\left(
\begin{array}{cc}
	U & V\\
	V^{\dagger} & W
\end{array}
\right)
\end{equation}
with $U\in H_m$ and $W\in H_k$.  In this basis we have
\begin{equation}
P	=\left(
\begin{array}{cc}
	I_m & 0\\
	0 & 0
\end{array}
\right),
\qquad\qquad
Q	=\left(
\begin{array}{cc}
	0 & 0\\
	0 & I_k
\end{array}
\right),
\end{equation}
and hence
\begin{equation}
{\mathbf{P}}X	=\left(
\begin{array}{cc}
	U & 0\\
	0 & 0
\end{array}
\right),
\qquad\qquad
{\mathbf{Q}}X	=\left(
\begin{array}{cc}
	0 & 0\\
	0 & W
\end{array}
\right),
\qquad\qquad
{\mathbf{R}}X	=\left(
\begin{array}{cc}
	0 & V\\
	V^{\dagger} & 0
\end{array}
\right).
\end{equation}


\section{Composite systems}
\label{sec:compositesystems}

In order to describe entanglement in quantum systems it is necessary
to develop the basic theory of tensor product spaces.  We do this for
a bipartite system consisting of two subsystems A and B.

\subsection{Product vectors}
\label{sec:productvectors}

If $N=N_AN_B$ then the tensor product spaces
${\mathbb{C}^N}={\mathbb{C}^{N_A}}\otimes{\mathbb{C}^{N_B}}$ (a
complex tensor product) and $H_N=H_{N_A}\otimes H_{N_B}$ (a real
tensor product) describe a composite quantum system with two
subsystems A and B of Hilbert space dimensions $N_A$ and $N_B$.

A vector $\psi\in{\mathbb{C}^N}$ then has components
$\psi_I=\psi_{ij}$, where
\begin{equation}
	I=1,\ldots,N
	\quad
	\leftrightarrow
	\quad
	ij=11,12,\ldots,1N_B,21,22,\ldots,N_AN_B\;.
\end{equation}
A product vector $\psi=\phi\otimes\chi$ has components
$\psi_{ij}=\phi_{i}\chi_{j}$. We see that $\psi$ is a product vector
if and only if its components satisfy the quadratic equations
\begin{equation}
	\label{productvectors}
	\psi_{ij}\psi_{kl}-\psi_{il}\psi_{kj}=0\;.
\end{equation}
These equations are not all independent, the number of independent
complex equations is
\begin{equation}
\label{eq:constrpv}
	K=(N_A-1)(N_B-1)=N-N_A-N_B+1\;.
\end{equation}
For example, if $\psi_1\neq 0$ we get a complete set of independent
equations by taking $i=j=1$, $k=2,\ldots,N_A$, and $l=2,\ldots,N_B$.

Since the Equations~\eqref{productvectors} are homogeneous, any
solution $\psi\neq 0$ gives rise to a one parameter family of
solutions $c\psi$ where $c\in {\mathbb{C}}$. A vector $\psi$ in a
subspace of dimension $n$ has $n$ independent complex
components. Since the most general nonzero solution must contain at
least one free complex parameter, we conclude that a generic subspace
of dimension $n$ will contain nonzero product vectors if and only if
\begin{equation}
	n\geq K+1\;.
\end{equation}
The limiting dimension
\begin{equation}
\label{eq:limdim}
	n=K+1=N-N_A-N_B+2
\end{equation}
is particularly interesting. In this special case a nonzero solution
will contain exactly one free parameter, which has to be a complex
normalization constant. Thus up to proportionality there will exist a
finite set of product vectors in a generic subspace of this dimension,
in fact the number of product vectors is~\cite{Hartshorne}
\begin{equation}
\label{eq:gennumpv}
	p=\left(
\begin{array}{c}
	N_A+N_B-2\\
	N_A-1
\end{array}
\right)
=\frac{(N_A+N_B-2)!}{(N_A-1)!(N_B-1)!}\;.
\end{equation}
A generic subspace of lower dimension will contain no nonzero product
vector, whereas any subspace of higher dimension will contain a
continuous infinity of different product vectors (different in the
sense that they are not proportional).

These results hold for \emph{generic} subspaces. It is trivially clear
that nongeneric subspaces with low dimensions exist that contain
product vectors.  In the special case $N_A=N_B=3$ studied here, the
limiting dimension given by Equation~\eqref{eq:limdim} is five, and
the number of product vectors given by Equation~\eqref{eq:gennumpv} is
six.  As described in Section \ref{sec:Averyspecialsubspace} we have
found one special example of a five dimensional subspace with only two
product vectors.  Nongeneric cases are treated more generally
in~\cite{Chen11}.

The facts that product vectors always exist in $K+1$ and higher
dimensions, but not always in $K$ and lower dimensions, are special
cases of a theorem proved by Parthasarathy for systems composed of any
number of subsystems~\cite{Parthasarathy04}.  These results have
profound implications for the construction of unextendible product
bases.

\subsection{Partial transposition and separability}

The following relation between matrix elements
\begin{equation}
	(X^P)_{ij;kl}=X_{il;kj}
\end{equation}
defines the partial transpose $X^P=X^{T_B}$ of the matrix $X$ with
respect to the second subsystem B.  The partial transpose with respect
to the first subsystem A is $X^{T_A}=X^{PT}$ where $T$ denotes total
transposition.  Since we work with Hermitian matrices, $T$ is the same
as complex conjugation. The partial transposition is transposition of
the individual submatrices of dimension $N_B\times N_B$ in the
$N\times N$ matrix $X$.  If $X=Y\otimes Z$ then $X^P=Y\otimes Z^T$.

A density matrix $\rho$ is called separable if it is a convex
combination of tensor product pure states,
\begin{equation}
\label{eq:sepdef}
  \rho=\sum_{k}\,p_k\,w_kw_k^{\,\dagger}\;,
\end{equation}
with $w_k=u_k\otimes v_k\in\mathbb{C}^N$, $p_k>0$, and
$\sum_k\,p_k=1$.  It follows that
\begin{equation}
  \rho^P
  =\sum_{k}\,p_k\,(u_ku_k^{\,\dagger})\otimes (v_kv_k^{\,\dagger})^T
  =\sum_{k}\,p_k\,(u_ku_k^{\,\dagger})\otimes (v_k^*v_k^{\,T})\;.
\end{equation}
We denote the set of separable density matrices by $\mathcal{S}$.

The obvious fact that $\rho^P$ is positive when $\rho$ is separable is
known as the Peres criterion, it is an easily testable necessary
condition for separability. For this reason it is of interest to study
the set of PPT or Positive Partial Transpose matrices, defined as
\begin{equation}
	\mathcal{P}=\{\,\rho\in{\mathcal{D}}\,|\,
             \rho^P \geq 0\,\}={\mathcal{D}} \cap {\mathcal{D}^P}\;.
\end{equation}
We may call it the Peres set. A well known result is that
$\mathcal{P}=\mathcal{S}$ for $N=N_AN_B\leq 6$, whereas $\mathcal{P}$
is strictly larger than $\mathcal{S}$ in higher dimensions
\cite{MPRHorodecki96}.

We will classify low rank PPT states by the ranks $(m,n)$ of $\rho$
and $\rho^P$ respectively.  Here we study the special case
$N_A=N_B=3$, then the ranks $(m,n)$ and $(n,m)$ are equivalent for the
purpose of classification, because of the symmetric roles of the
subsystems A and B, and the arbitrariness of choice of which subsystem
to partial transpose.

\subsection{Product vectors in the kernel and range}
\label{prodvecker}

Recall that ${\psi}^{\dagger}\rho{\psi}=0\Leftrightarrow \rho\psi=0$
when $\rho\geq 0$, and similarly for $\rho^P$.  The identity
\begin{equation}
  (x\otimes y)^{\dagger}\rho(x\otimes y)=
  (x\otimes y^*)^{\dagger}\rho^P(x\otimes y^*)
\end{equation}
therefore implies, for a PPT state $\rho$, that
$x\otimes y\in\textrm{Ker}\,\rho$ if and only if
$x\otimes y^*\in\textrm{Ker}\,\rho^P$.

Let the number of product vectors in $\textrm{Img}\,\rho$,
$\textrm{Img}\,\rho^P$ and $\textrm{Ker}\,\rho$ be respectively
$n_{\textrm{img}}$, $\widetilde{n}_{\textrm{img}}$, and
$n_{\textrm{ker}}$.  Then $n_{\textrm{ker}}$ is also the number of
product vectors in $\textrm{Ker}\,\rho^P$, and $\rho$ is characterized
by
$\{n_{\textrm{img}},\widetilde{n}_{\textrm{img}};n_{\textrm{ker}}\}$.
We find numerically that the generic entangled PPT states of rank
$(5,5)$ are by this characterization $\{6,6;0\}$ states.

We write the product vectors in $\textrm{Img}\,\rho$ as
\begin{equation}
	w_i = u_i\otimes v_i\;, \qquad i=1,\ldots,n_{\textrm{img}}\;.
\end{equation}
And likewise for $\textrm{Ker}\,\rho$,
\begin{equation}
	z_j = x_j\otimes y_j\;, \qquad j=1,\ldots,n_{\textrm{ker}}\;.
\end{equation}
Since the two subspaces are orthogonal, it is necessary that
$w_i^{\dagger}z_j=0$ for all $i,j$, hence for every pair $i,j$ we must
have either $u_i^{\dagger}x_j=0$ or $v_i^{\dagger}y_j=0$.

\subsection{The range criterion and edge states}
\label{subsubsecrangecrit}

If $\rho$ is the separable state given in Equation~\eqref{eq:sepdef},
then $\textrm{Img}\,\rho$ is spanned by the product vectors
$w_k=u_k\otimes v_k$, and $\textrm{Img}\,\rho^P$ is spanned by the
partially conjugated product vectors
$\widetilde{w}_k=u_k\otimes v_k^*$.  No such relation is known to
exist between product vectors in $\textrm{Img}\,\rho$ and
$\textrm{Img}\,\rho^P$ when $\rho$ is an entangled PPT state.

The existence of a set of product vectors spanning the range of
$\rho$, such that the partially conjugated product vectors span the
range of $\rho^P$, is therefore seen to be a necessary condition for
separability, called the \emph{range criterion} \cite{PHorodecki97}.
Since there exist entangled states satisfying the range criterion, the
condition is not sufficient.

Please note that the range criterion demands that there should exist
at least one such set of product vectors, not that all product vectors
in the ranges should be related by partial conjugation.  For example,
in $3\times 3$ dimensions, which is the case discussed here, when a
separable state $\rho$ is a convex combination of five randomly chosen
pure product states, $\textrm{Img}\,\rho$ and $\textrm{Img}\,\rho^P$
will contain exactly six product vectors each, but the sixth product
vector in $\textrm{Img}\,\rho^P$ is not the partial conjugate of the
sixth product vector in $\textrm{Img}\,\rho$.

An \emph{edge state} is defined as a PPT state that breaks the range
criterion maximally, in the sense that there exists no product vector
in the range of $\rho$ with its partial conjugate in the range of
$\rho^P$.  All edge states are entangled by the range criterion.

It is straightforward to see that every extremal entangled PPT state
$\rho$ must be an edge state.  In fact, if
$w=u\otimes v\in\textrm{Img}\,\rho$ and
$\widetilde{w}=u\otimes v^*\in\textrm{Img}\,\rho^P$, then $\rho$ is
not extremal, because $(1-\epsilon)\rho+\epsilon\,ww^{\dagger}$ is a
PPT state for both positive and negative $\epsilon$ in some finite
interval.  The converse is not true: there exist plenty of edge states
that are not extremal~\cite{Leinaas10B}.

\subsection{Extremality in $\mathcal{P}$}
\label{subsec:extremalitytest}

We will now describe the extremality test in
$\mathcal{P}=\mathcal{D}\cap\mathcal{D}^P$, which follows directly
from the extremality test in $\mathcal{D}$.  We want to outline also
how to use perturbations with various restrictions in order to
calculate the dimensions of surfaces of states in $\mathcal{P}$ of
fixed ranks.  Thus we are interested in perturbations that preserve
the ranks $(m,n)$ of $\rho$ and $\rho^P$ simultaneously, but do not
necessarily preserve the ranges.

As we did for $\rho$, we define
$\widetilde{P}$ and $\widetilde{Q}=I-\widetilde{P}$
as the orthogonal projections
onto $\textrm{Img}\,\rho^P$ and $\textrm{Ker}\,\rho^P$.  We then
define
\begin{eqnarray}
{\mathbf{\widetilde{P}}}X & \!\!\! = & \!\!\!
(\widetilde{P}X^P\widetilde{P})^P\,,\nonumber\\ 
{\mathbf{\widetilde{Q}}}X & \!\!\! = & \!\!\!
(\widetilde{Q}X^P\widetilde{Q})^P\,,\nonumber\\
{\mathbf{\widetilde{R}}}X & \!\!\! = & \!\!\!  
(\mathbf{I-\widetilde{P}-\widetilde{Q}})X\;.
\end{eqnarray}
These are again projections on the real Hilbert space $H_N$, and like
${\mathbf{P}},\,{\mathbf{Q}}$ and ${\mathbf{R}}$ they are symmetric
with respect to the natural scalar product on $H_N$. We use these
projection operators on $H_N$ to impose various restrictions on the
perturbation matrix $A$ in Equation~\eqref{eq:pertrho}.

\subsection*{Testing for extremality in $\mathcal{P}$}

Clearly $\rho$ is extremal in $\mathcal{P}$ if and only if $A=\rho$ is
the only simultaneous solution of the two equations ${\mathbf{P}}A=A$
and ${\mathbf{\widetilde{P}}}A=A$.  Another way to formulate this
condition is that there exists no
$\rho\,'\in\mathcal{P},\;\rho\,'\neq\rho$, with both
$\textrm{Img}\,\rho\,'=\textrm{Img}\,\rho$ and
$\textrm{Img}\,(\rho\,')^P=\textrm{Img}\,\rho^P.$
 
Since ${\mathbf{P}}$ and ${\mathbf{\widetilde{P}}}$ are projections,
the equations ${\mathbf{P}}A=A$ and ${\mathbf{\widetilde{P}}}A=A$
together are equivalent to the single eigenvalue equation
\begin{equation}
	\label{extremality}
	({\mathbf{P}}+{\mathbf{\widetilde{P}}})A=2A\;.
\end{equation}
Note that the operator ${\mathbf{P}}+{\mathbf{\widetilde{P}}}$
is real symmetric and positive and therefore has a complete set of
nonnegative real eigenvalues and eigenvectors.

When we diagonalize ${\mathbf{P}}+{\mathbf{\widetilde{P}}}$ we will
always find $A=\rho$ as an eigenvector with eigenvalue 2. If it is the
only solution of Equation~\eqref{extremality}, this proves that $\rho$ is
extremal in $\mathcal{P}$. If $A$ is a solution not proportional to
$\rho$, then we may impose the condition $\textrm{Tr}\,A=0$ (replace
$A$ by $A-(\textrm{Tr}A)\rho$ if necessary), and we know that there
exists a finite range of both positive and negative values of
$\epsilon$ such that $\rho\,'=\rho+\epsilon A\in\mathcal{P}$, hence
$\rho$ is not extremal in $\mathcal{P}$.

\subsection*{Perturbations preserving the PPT property and ranks}

The rank and positivity of $\rho$ is preserved by the perturbation
$\rho\,'=\rho+\epsilon A$ to first order in $\epsilon$, both for
$\epsilon>0$ and $\epsilon<0$, if and only if
$\mathbf{Q}A=0$. Similarly, the rank and positivity of $\rho^P$ is
preserved if and only if $\mathbf{\widetilde{Q}}A=0$. These two
equations together are equivalent to the single eigenvalue equation
\begin{equation}
	\label{QQA}
	({\mathbf{Q}}+{\mathbf{\widetilde{Q}}})A=0\;.
\end{equation}
Again ${\mathbf{Q}}+{\mathbf{\widetilde{Q}}}$ is real symmetric and
has a complete set of real eigenvalues and eigenvectors.  The number
of linearly independent solutions for $A$ in Equation~\eqref{QQA} is
then the dimension of the surface through $\rho$ of rank $(m,n)$ PPT
states.

We may want to perturb in such a way that
$\textrm{Img}\,\rho\,'=\textrm{Img}\,\rho$, but not necessarily
$\textrm{Img}\,(\rho\,')^P=\textrm{Img}\,\rho^P$, that is, we only
require $\textrm{Img}\,(\rho\,')^P$ and $\textrm{Img}\,\rho^P$ to have
the same rank. Then the conditions on $A$ are that ${\mathbf{P}}A=A$
and $\mathbf{\widetilde{Q}}A=0$, or equivalently
\begin{equation}
	\label{IPQ}
	({\mathbf{I}}-{\mathbf{P}}+{\mathbf{\widetilde{Q}}})A=0\;.
\end{equation}
In this case the number of linearly independent solutions for $A$ is
the dimension of the surface through $\rho$ of rank $(m,n)$ PPT states
with fixed range $\textrm{Img}\,\rho$.

\subsection{Product transformations}
\label{subsec:producttrans}

A product transformation of the form
\begin{equation}
	\label{SL-transformation1}
	\rho\mapsto\rho\,'=a\,V\!\rho V^{\dagger}\;,
	\qquad\qquad
	V=V_A\otimes V_B
\end{equation}
where $a>0$ is a normalization factor and
$V_A\in \textrm{SL}(N_A,\mathbb{C})$,
$V_B\in \textrm{SL}(N_B,\mathbb{C})$, preserves positivity, rank,
separability and other interesting properties that the density matrix
$\rho$ may have. It preserves positivity of the partial transpose
because
\begin{equation}
	(\rho\,')^P=a\,\widetilde{V}\!\rho^P {\widetilde{V}}^{\dagger}\;,
	\qquad
	\qquad
	\widetilde{V}=V_A\otimes {V_B}^{*}\;.
\end{equation}
A transformation of the form of Equation~\eqref{SL-transformation1} is also
sometimes referred to as a \emph{local} SL-transformation.

The range and kernel of $\rho$ and $\rho^P$ transform in the following
ways,
\begin{eqnarray}
\textrm{Img}\,\rho\,'=&\!\!\!V\,\textrm{Img}\,\rho\;,
\qquad\qquad \qquad 
\textrm{Ker}\,\rho\,'&\!\!\!=(V^{\dagger})^{-1}\,\textrm{Ker}\,\rho\;,
\nonumber\\
\textrm{Img}\,(\rho\,')^P=&\!\!\!\widetilde{V}\,\textrm{Img}\,\rho^P\,,
\qquad\qquad 
\textrm{Ker}\,({\rho\,'})^P&\!\!\!=
({\widetilde{V}}^{\dagger})^{-1}\,\textrm{Ker}\,\rho^P\,.
\end{eqnarray}
We say that two density matrices $\rho$ and $\rho\,'$ related in this
way are $\textrm{SL}\otimes\textrm{SL}$-equivalent. The concept of
$\textrm{SL}\otimes\textrm{SL}$-equivalence is important
because it simplifies very much the classification of the low rank PPT
states. Since this $\textrm{SL}\otimes\textrm{SL}$-equivalence is transitive
it generates equivalence classes of matrices.

\subsubsection*{$\textrm{SL}\otimes\textrm{SL}$-symmetry under partial transposition}

We say that the
state $\rho$ is $\textrm{SL}\otimes\textrm{SL}$-symmetric
if $\rho$ and $\rho^P$ are $\textrm{SL}\otimes\textrm{SL}$-equivalent, that is, if
\begin{equation}
	\label{SL-symmetric}
	\rho^P=a\,V\!\rho V^{\dagger}\;,
	\qquad\qquad
	V=V_A\otimes V_B\;.
\end{equation}

Since SL-transformations of product type $V=V_A\otimes V_B$ preserve
the number of product vectors in a subspace, any transformation
$\rho\mapsto\rho^P=a\,V\!\rho V^{\dagger}$ must transform the set of
$n_{\textrm{img}}$ product vectors in the range of $\rho$ to the set
of $\widetilde{n}_{\textrm{img}}$ product vectors in the range of
$\rho^P$, so for $\textrm{SL}\otimes\textrm{SL}$-symmetric states we
must have $n_{\textrm{img}}=\widetilde{n}_{\textrm{img}}$. If the
product vectors in $\textrm{Img}\,\rho$ are $w_i=u_i\otimes v_i$ with
$i=1,\ldots,n_{\textrm{img}}$ and the product vectors in
$\textrm{Img}\,\rho^P$ are
$\widetilde{w}_i=\widetilde{u}_i\otimes \widetilde{v}_i$ for
$i=1,\ldots,n_{\textrm{img}}$, then
\begin{equation}
	\label{VAVB}
	V_A\,u_i=\widetilde{u}_i
	\qquad
	V_B\,v_i=\widetilde{v}_i.
\end{equation}
Since our
understanding of the relation between the sets $\{w_i\}$ and
$\{\widetilde{w}_i\}$ is quite limited for entangled states, it is
difficult to say much in general about what makes some states
$\textrm{SL}\otimes\textrm{SL}$-symmetric and others not.

\subsubsection*{Genuine $\textrm{SL}\otimes\textrm{SL}$-symmetry}

We say that a state $\rho$ is
\emph{genuinely} $\textrm{SL}\otimes\textrm{SL}$-symmetric if there
exists a transformation
\begin{equation}
  \label{eq:genuineSLxSL}
  \rho'=a\,U\!\rho U^{\dagger}\;,
  \qquad
  \qquad
  U=U_A\otimes U_B
\end{equation}
such that $(\rho')^P=\rho'$.  The transformation of $\rho$ implies
that
\vspace*{-0.2cm}
\begin{equation}
(\rho')^P=a\,\widetilde{U}\!\rho^P\widetilde{U}^{\dagger}
\end{equation}
when we define $\widetilde{U}=U_A\otimes U_B^*$.  The genuine
$\textrm{SL}\otimes\textrm{SL}$-symmetry implies further that
\begin{equation}
\widetilde{U}\!\rho^P\widetilde{U}^{\dagger}=U\!\rho U^{\dagger}\;,
\end{equation}
and hence
\begin{equation}
\label{eq:rhoPVrhoV}
  \rho^P=\,V\!\rho V^{\dagger}\qquad\mathrm{with}\qquad
  V=\widetilde{U}^{-1}U=I\otimes V_B\;,
\end{equation}
and with $V_B=(U_B^*)^{-1}U_B$.  This shows that genuine
$\textrm{SL}\otimes\textrm{SL}$-symmetry
implies $\textrm{SL}\otimes\textrm{SL}$-symmetry, and
Equation~\eqref{eq:rhoPVrhoV} requires that $V$
preserves the trace of $\rho$,
\begin{equation}
\textrm{Tr}\,\rho=\textrm{Tr}\,\rho^P=
\textrm{Tr}\,(V\rho V^{\dagger})=\textrm{Tr}\,(\rho V^{\dagger}V)\;.
\end{equation}
A sufficient but not necessary condition for this trace preservation
is that $V_B$ is unitary.

The relation $V_B=(U_B^*)^{-1}U_B$ implies that $V_B$ has some special
properties.  One implication is that $V_B^*=V_B^{\,-1}$, and hence
$|\det V_B|=1$.
  If we multiply $U_B$ by some phase factor
$\textrm{e}^{\textrm{i}\alpha}$, then $V_B$ is multiplied by
$\textrm{e}^{2\textrm{i}\alpha}$, and in this way we may redefine
$V_B$ such that $\det V_B=1$.

We conclude that for the state $\rho$ to be genuinely
$\textrm{SL}\otimes\textrm{SL}$-symmetric it must be
$\textrm{SL}\otimes\textrm{SL}$-symmetric with a
transformation of the form given in Equation~\eqref{eq:rhoPVrhoV}.  For the
$3\times 3$ system, $V$ would have the form
\begin{equation}
	\label{genuineSL}
	V=
\begin{pmatrix}
	V_B & 0 & 0\\
	0 & V_B & 0\\
	0 & 0 & V_B
\end{pmatrix}
\end{equation}
with $V_B\in\textrm{SL}(3,\mathbb{C})$ and $V_B^*=V_B^{\,-1}$.

Since the transformation has the form $V=I\otimes V_B$ in the case of
genuine $\textrm{SL}\otimes\textrm{SL}$-symmetry, by Equation~\eqref{VAVB} the product
vectors in $\textrm{Img}\,\rho$ and $\textrm{Img}\,\rho^P$ will be
related by the transformations $\widetilde{u}_i=u_i$ and
$\widetilde{v}_i=V_Bv_i$.  This is a necessary condition for genuine
$\textrm{SL}\otimes\textrm{SL}$-symmetry which may be tested as soon as
we know the product vectors
in $\textrm{Img}\,\rho$ and $\textrm{Img}\,\rho^P$.

Assume that for a given PPT state $\rho$ we find that $\rho$ and
$\rho^P$ are related by a transformation of the form given in
Equation~\eqref{eq:rhoPVrhoV}.  Then a further problem to be solved is
to find a transformation $U=U_A\otimes U_B$ that demonstrates
explicitly the genuine $\textrm{SL}\otimes\textrm{SL}$-symmetry of
$\rho$.  Here $U_A$ is completely arbitrary, hence the simplest
solution is to take $U_A=I$.  Next we have to solve the equation
$V_B=(U_B^*)^{-1}U_B$ for $U_B$.  We find one particular solution by
assuming that $U_B^*=U_B^{\,-1}$.  This gives the equation
$V_B=(U_B)^2$, and the solution is the matrix square root,
\begin{equation}
\label{eq:IsqrtVB}
U_B=\sqrt{V_B}\;.
\end{equation}
Since $V_B^*=V_B^{\,-1}$, this is consistent with the assumption
$U_B^*=U_B^{\,-1}$.

It should be noted that the matrix square root is in general many
valued.  A standard solution $U_B=X$ is found by the rapidly
converging Newton--Raphson method: starting with $X=V_B$ we iterate
the substitution
\begin{equation}
X\to X'=\frac{1}{2}\,(X+X^{-1}V_B)\;.
\end{equation}
We find in practice that this solution for $U=I\otimes U_B$ always
works.


\section{The rank $(4,4)$ extremal PPT states revisited}
\label{sec:rank44revisited}

The rank $(4,4)$ extremal PPT states in dimension $3\times 3$ are well
understood \cite{Leinaas10A,Chen11,Skowronek11,Hansen12}.  They are all
$\textrm{SL}\otimes\textrm{SL}$-equivalent to states constructed from
unextendible product bases (UPBs), and they all have exactly six
product vectors in their kernels.  A construction method not using
UPBs was discussed in~\cite{Hansen12}, and as part of that discussion
the structure of a PPT state $\rho$ with at least four product vectors
in $\textrm{Ker}\,\rho$ was derived.

In the present section we will review and expand on the discussion
given in~\cite{Hansen12}.  This is relevant for our present study of
nongeneric rank $(5,5)$ PPT states, and it leads to a new construction
of the rank $(4,4)$ extremal PPT states.

Given a PPT state $\rho$ of rank at most five, and four product
vectors $z_j=x_j\otimes y_j$ in $\textrm{Ker}\,\rho$, in some definite
but arbitrary order.  We assume that any three $x$ vectors and any
three $y$ vectors are linearly independent.  Then we may perform a
product transformation as in Equation~\eqref{SL-transformation1}, and
subsequent normalizations, so that the vectors take the form
\begin{equation}
\label{eq:kerxy}
x=y=
\left(
\begin{array}{cccc}
	1 & 0 & 0 & 1\\
	0 & 1 & 0 & 1\\
	0 & 0 & 1 & 1
\end{array}
\right)
\end{equation}
and
\begin{equation}
\label{eq:kerz}
z=
\left(
\begin{array}{cccc}
	1 & 0 & 0 & 1\\
	0 & 0 & 0 & 1\\
	0 & 0 & 0 & 1\\
	0 & 0 & 0 & 1\\
	0 & 1 & 0 & 1\\
	0 & 0 & 0 & 1\\
	0 & 0 & 0 & 1\\
	0 & 0 & 0 & 1\\
	0 & 0 & 1 & 1\\
\end{array}
\right)\;.
\end{equation}
The transformation is unique.  In this four dimensional subspace there
exist no other product vectors.  The real form of the product vectors
$z_j\in\mathrm{Ker}\,\rho$ implies that $z_j\in\mathrm{Ker}\,\rho^P$.

It is equally easy to see that there exist exactly six product vectors
$w_i = u_i\otimes v_i$ in the orthogonal subspace.  In fact, in order
to have $(u_i\otimes v_i)\perp(x_j\otimes y_j)$ for all $i=1,\ldots,6$
and $j=1,\ldots,4$, we must have for each pair $i,j$ that either
$u_i\perp x_j$ or $v_i\perp y_j$.  Since any three $x$ vectors and any
three $y$ vectors are linearly independent, a $u$ vector can be
orthogonal to at most two $x$ vectors, and a $v$ vector can be
orthogonal to at most two $y$ vectors.  This gives the six
possibilities for orthogonality listed in the table.
\begin{table}[H]
\begin{center}
\begin{tabular}{|c|c|c|}
\hline
$u_i\otimes v_i$ & $u_i\perp x_k,x_l$ & $v_i\perp y_m,y_n$\\
\hline
$i$ & $k,l$ & $m,n$\\
\hline
$1$ & $2,3$ & $1,4$\\
$2$ & $1,3$ & $2,4$\\
$3$ & $1,2$ & $3,4$\\
$4$ & $1,4$ & $2,3$\\
$5$ & $2,4$ & $1,3$\\
$6$ & $3,4$ & $1,2$\\
\hline
\end{tabular} 
\end{center}
\caption{The six possible ways to have a product vector
$u_i\otimes v_i$ orthogonal to all four product vectors
$x_j\otimes y_j$.}
\end{table}
The unique solution is the following:
\begin{equation}
\label{eq:64standard}
u=
\left(\!\!\!\!
\begin{array}{rrrrrr}
	\phantom{-}1 &  0 &  0 &  0 &  1 &  1\\
	0 &  \phantom{-}1 &  0 &  1 &  0 & -1\\
	0 &  0 & \phantom{-}1 & -1 & -1 &  0    
\end{array}
\right)\;,
\qquad
v=
\left(\!\!
\begin{array}{rrrrrr}
	 0 &  1 &  1 &  \phantom{-}1 &  0 &  0\\
	 1 &  0 & -1 &  0 &  \phantom{-}1 &  0\\
	-1 & -1 &  0 &  0 &  0 & \phantom{-}1    
\end{array}
\right)\;,
\end{equation}
\begin{equation}
\label{eq:64}
w=
\left(
\begin{array}{rrrrrr}
	0 & 0 & 0 & 0 & 0 & 0\\
	1 & 0 & 0 & 0 & 1 & 0\\
       -1 & 0 & 0 & 0 & 0 & 1\\
	0 & 1 & 0 & 1 & 0 & 0\\
	0 & 0 & 0 & 0 & 0 & 0\\
	0 &-1 & 0 & 0 & 0 &-1\\
	0 & 0 & 1 &-1 & 0 & 0\\
	0 & 0 &-1 & 0 &-1 & 0\\
	0 & 0 & 0 & 0 & 0 & 0
\end{array}
\right).
\end{equation} 

The five dimensional subspace $\textrm{Span}\,w$ given
by~\eqref{eq:64} defines a face $\mathcal{F}\subset\mathcal{D}$ of
dimension $5^2-1=24$.  Recall that
\begin{equation}
\label{eq:160317}
\rho z_j=\rho^P z_j=0\;,\qquad  j=1,2,3,4\;.
\end{equation} 
In terms of the face $\mathcal{F}$, Equation~\eqref{eq:160317} means
that $\rho\in\mathcal{F}$ and $\rho^P\in\mathcal{F}$, or equivalently,
that $\rho\in\mathcal{F}\cap\mathcal{F}^P$.
Equation~\eqref{eq:160317} restricts $\rho$ to have the following
form, with real coefficients $c_i$,
\begin{equation}
\settowidth{\mycolwd}{$c_1+c_5$}
\label{separablestate664}
\rho=
\frac{1}{2}\left(
\begin{array}{ccccccccc}
\phantom{-}0\phantom{-} & 0       & 0       & 0       &\phantom{-}0\phantom{-} & 0       & 0       & 0      &\phantom{-}0\phantom{-}\\
0 & c_1+c_5 & -c_1
    & 0       & 0 & 0       & 0       &-c_5     & 0\\
0 &-c_1     & c_1+c_6 & 0       & 0 &-c_6     & 0       & 0       & 0\\
0 & 0       & 0       & c_2+c_4 & 0 &-c_2     &-c_4     & 0       & 0\\
0 & 0       & 0       & 0       & 0 & 0       & 0       & 0       & 0\\
0 & 0       & -c_6    & -c_2    & 0 & c_2+c_6 & 0       & 0       & 0\\
0 & 0       & 0       & -c_4    & 0 & 0       & c_3+c_4 &-c_3     & 0\\
0 &-c_5     & 0       & 0       & 0 & 0       &-c_3     & c_3+c_5 & 0\\
0 & 0       & 0       & 0       & 0 & 0       & 0       & 0       & 0
\end{array}
\right).
\end{equation}
This form implies that $\rho^P=\rho$.  Since $\mathcal{F}$ is a face
on $\mathcal{D}$, we see that $\mathcal{F}^P$ is a face on $\mathcal{D}^P$, and
the intersection $\mathcal{G}=\mathcal{F}\cap\mathcal{F}^P$ is a face
on $\mathcal{P}=\mathcal{D}\cap\mathcal{D}^P$.
Equation~\eqref{separablestate664} with the normalization condition
\begin{equation}
\label{eq:ciwiwidaggernorm}
\textrm{Tr}\,\rho=\sum_{i=1}^6 c_i=1
\end{equation}
shows that the face $\mathcal{G}$ has dimension five.

The state $\rho$ defined in Equation~\eqref{separablestate664} is a
linear combination, but not necessarily a convex combination, of the
six pure product states,
\begin{equation}
\label{eq:ciwiwidagger}
\rho=\frac{1}{2}\,\sum_{i=1}^6 c_i\,w_iw_i^{\dagger}\;.
\end{equation}
Since the matrices $w_iw_i^{\dagger}$ are linearly independent, and
there are no other pure product states in $\textrm{Img}\,\rho$, we see that $\rho$
is separable if and only if all the coefficients $c_i$ are
nonnegative.  However, we will now see that it is possible for $\rho$
to be an entangled PPT state even if one of the coefficients is
negative.

An eigenvalue of $\rho$, and of $\rho^P=\rho$, is a root of the
characteristic polynomial
\begin{equation}
  \det(\rho-\lambda I)
  =-\lambda^4\,(\lambda^5-d_4\lambda^4+d_3\lambda^3
  -d_2\lambda^2+d_1\lambda-d_0)\;,
\end{equation}	
with
\begin{equation}
d_0
=\frac{3}{16}\,\sum_{i=1}^6\prod_{j\neq i}c_j
=\frac{3}{16}\,\Big(\prod_{j=1}^6c_j\Big)\sum_{i=1}^6 \frac{1}{c_i}\;.
\end{equation}	

If we start with $c_i>0$ for $i=1,\ldots,6$, then $\rho$ is a rank
$(5,5)$ separable state.  If we next change the coefficients
continuously, $\rho$ will continue to have five positive eigenvalues
until we get $d_0=0$.  Hence the equation $d_0=0$ defines the boundary
of the set of density matrices, and also of the set of PPT states
since $\rho^P=\rho$.

We know that the boundary is not reached before at least one
coefficient $c_i$ becomes zero or negative.  If two coefficients
become zero simultaneously, then $d_0=0$ and we have reached a
boundary state which is separable.  To get negative coefficients while
$\rho$ is a rank $(5,5)$ PPT state we have to make one coefficient
negative before the others.  Let us say, for example, that $c_1<0$,
and that we want to make also $c_2$ negative, while $c_i>0$ for
$i=3,4,5,6$.  Then we first have to make $c_2=0$, in which case
$d_0=3\,c_1c_3c_4c_5c_6/16<0$ and we have already crossed the boundary
$d_0=0$.

In conclusion, the entangled boundary states have $c_i\neq 0$ for
$i=1,\ldots,6$, and they have one negative and five positive
coefficients $c_i$ satisfying the equation
\begin{equation}
\label{eq:44PPT160211}
\sum_{i=1}^6 \frac{1}{c_i}=0\;.
\end{equation}	
Thus the boundary $d_0=0$ consists of two types of states.
\begin{enumerate}

\item
  Separable states that are convex combinations of up to four of the
  pure product states $w_iw_i^{\,\dagger}$.

\item
  Rank $(4,4)$ entangled PPT states that are linear combinations of
  all the six pure product states $w_iw_i^{\,\dagger}$ with exactly
  one negative coefficient.

\end{enumerate}
It is well known that rank $(4,4)$ entangled PPT states are extremal.
\begin{figure}[H]
\begin{center}
\includegraphics[width=16cm]{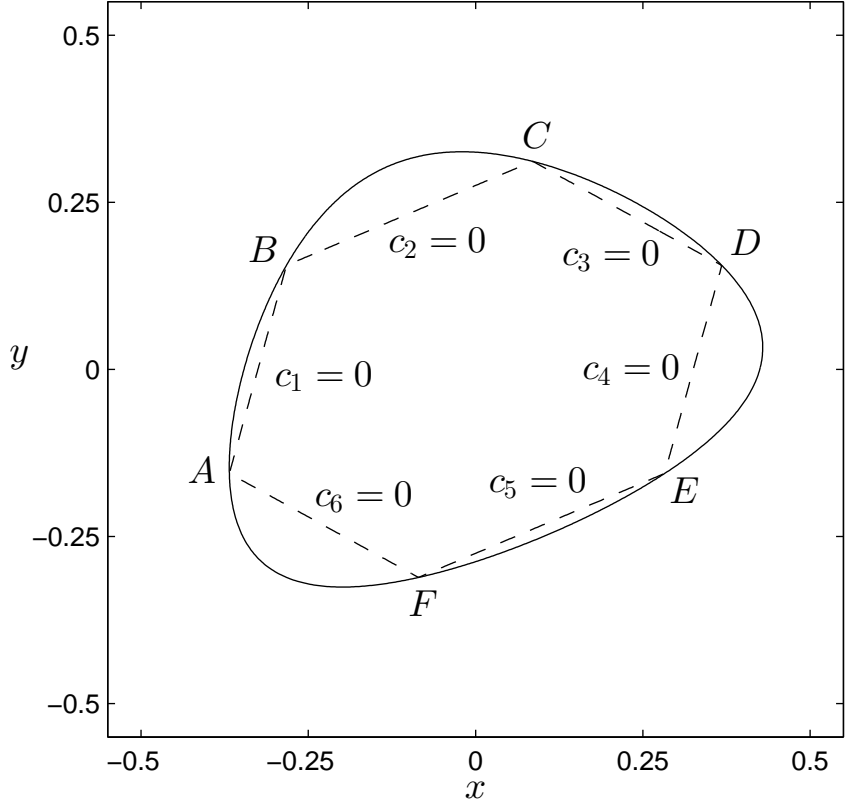}
\caption{\label{fig:plot080216} A two dimensional section through the
  five dimensional face on the set of PPT states consisting of
  normalized density matrices of the form in
  Equation~\eqref{separablestate664}.  The outer curve is the common
  boundary of $\mathcal{D}$ and $\mathcal{P}$, it consists of rank
  $(4,4)$ PPT states.  The hexagon (dashed) is the boundary of the
  simplex of separable states, where one of the coefficients $c_i$ is
  zero.  The region between the two curves consists of entangled PPT
  states of rank $(5,5)$ with one coefficient negative.  The
  coefficients $c_i$ defining the separable boundary states $A$ to $F$
  are given in Table~\ref{tab:coeffAF}.}
\end{center}
\end{figure}
Figure~\ref{fig:plot080216} shows an example of a two dimensional
section through the five dimensional face of $\mathcal{P}$ defined by
Equations~\eqref{eq:ciwiwidagger} and \eqref{eq:ciwiwidaggernorm}.
The section through $\mathcal{S}$ is the hexagon with corners $A$ to
$F$.  In Table~\ref{tab:coeffAF} we have listed the coefficients
$c_i$, multiplied by 12, that define the states $A$ to $F$ by
Equation~\eqref{eq:ciwiwidagger}.  The hexagon is reflection symmetric
about two axes.
\begin{table}[H]
\begin{center}
\begin{tabular}{|c|cccccc|}
\hline
$i$ & $A$ & $B$ & $C$ & $D$ & $E$ & $F$\\
\hline
$1$ & 0 & 0 & 3 & 6 & 6 & 3\\
$2$ & 1 & 0 & 0 & 1 & 2 & 2\\
$3$ & 6 & 3 & 0 & 0 & 3 & 6\\
$4$ & 2 & 2 & 1 & 0 & 0 & 1\\
$5$ & 3 & 6 & 6 & 3 & 0 & 0\\
$6$ & 0 & 1 & 2 & 2 & 1 & 0\\
\hline
\end{tabular} 
\end{center}
\caption{\label{tab:coeffAF}The coefficients $c_i$, multiplied by 12,
  for the states $A$ to $F$ in Figure~\ref{fig:plot080216}.}
\end{table}

\subsection*{The most general rank $(4,4)$ entangled PPT states}

Consider a general rank $(4,4)$ entangled PPT state $\rho$ in
$3\times 3$ dimensions.  It is known that any such state is extremal,
and has exactly six product vectors in its kernel.  We can now see
that it is $\textrm{SL}\otimes\textrm{SL}$-equivalent, in no less
than 360 different ways, to such
states on the boundary of the five dimensional face of $\mathcal{P}$
that we have described here.  The 360 transformations are found in the
following way.

Pick any four of the six product vectors in $\textrm{Ker}\,\rho$, this
can be done in 15 different ways.  Order them next in one of the 24
possible ways.  Altogether there are $24\times 15=360$ possibilities.
There is then a unique product transformation that transforms the four
product vectors to the form given in Equation~\eqref{eq:kerz}.  We know that
it must transform the state $\rho$ into one of the rank $(4,4)$ states
described by the Equations~\eqref{eq:ciwiwidagger},
\eqref{eq:ciwiwidaggernorm}, and~\eqref{eq:44PPT160211}, since these
are the only rank $(4,4)$ entangled PPT states of this form.


\section{Rank $(5,5)$ PPT states in $3 \times 3$ dimensions}
\label{sec:generic55}

Our main purpose with the present study has been to try to understand
the rank $(5,5)$ entangled PPT states in $3 \times 3$ dimensions.  In
particular, we would like to understand better the relation between
$\textrm{Img}\,\rho$ and $\textrm{Img}\,\rho^P$ when $\rho$ is a rank
$(5,5)$ PPT state.  A natural question is whether $\rho$ is
$\textrm{SL}\otimes\textrm{SL}$-symmetric, as defined in
Equation~\eqref{SL-symmetric}, so that
$\textrm{Img}\,\rho^P=V\,\textrm{Img}\,\rho$ with $V=V_A\otimes V_B$.

The product vectors in $\textrm{Img}\,\rho$ and $\textrm{Img}\,\rho^P$
are very useful for answering the question of
$\textrm{SL}\otimes\textrm{SL}$-symmetry, especially when both these
spaces have dimension five.  By Equation~\eqref{eq:constrpv}, the
number of constraints to be satisfied by a product vector is $K=4$,
thus $K+1=5$ is precisely the critical dimension at which every
subspace contains one or more product vectors, and a generic subspace
contains a finite number of product vectors, exactly six in this case.
{From} these product vectors one may construct invariants that may be
used to test whether the two spaces are related by some product
transformation $V=V_A\otimes V_B$, which is a necessary condition for
the $\textrm{SL}\otimes\textrm{SL}$-symmetry of the state $\rho$.

\subsection{Generic five dimensional subspaces}

Any set of five product vectors $w_i=u_i\otimes v_i$ in a generic five
dimensional subspace may be transformed by an
$\textrm{SL}\otimes\textrm{SL}$-transformation, followed by suitable
normalizations, to the standard form~\cite{Hansen12}
\begin{equation}
\label{sformpqrs}
u=
\left(
\begin{array}{ccccc}
	1 & 0 & 0 & 1 & 1\\
	0 & 1 & 0 & 1 & p\\
	0 & 0 & 1 & 1 & q
\end{array}
\right),
\qquad
v=
\left(
\begin{array}{ccccc}
	1 & 0 & 0 & 1 & 1\\
	0 & 1 & 0 & 1 & r\\
	0 & 0 & 1 & 1 & s
\end{array}
\right),
\end{equation}
with $p,q,r,s$ as real or complex parameters. By generic we here mean
that any three vectors in $u$ and in $v$ are linearly independent.
There will also be a sixth product vector $w_6=u_6\otimes v_6$ which
is a linear combination of the above five. The parameters $p,q,r,s$
are determined by the following ratios of determinants,
\begin{align}
	s_1 = & -\frac{\textrm{det}(u_1u_2u_4)\,\textrm{det}(u_1u_3u_5)}{\textrm{det}(u_1u_2u_5)\,\textrm{det}(u_1u_3u_4)}=-\frac{p}{q}\;,
\nonumber\\
	\label{s1s2}
	s_2 = & -\frac{\textrm{det}(u_1u_2u_3)\,\textrm{det}(u_2u_4u_5)}{\textrm{det}(u_1u_2u_4)\,\textrm{det}(u_2u_3u_5)}=q-1\;,
\end{align}

\begin{align}
	s_3 = & \,\,\frac{\textrm{det}(v_1v_2v_3)\,\textrm{det}(v_1v_4v_5)}{\textrm{det}(v_1v_2v_5)\,\textrm{det}(v_1v_3v_4)}=\frac{r-s}{s}\;, \nonumber \\
	\label{s3s4}
	s_4 = & \,\,\frac{\textrm{det}(v_1v_3v_5)\,\textrm{det}(v_2v_3v_4)}{\textrm{det}(v_1v_2v_3)\,\textrm{det}(v_3v_4v_5)}=\frac{r}{1-r}\;.
\end{align}
All the parameters $s_i$ are invariant under
$\textrm{SL}\otimes\textrm{SL}$-transformations.  Hence for given
vectors $u_i$ and $v_i$ not on standard form these formulas may be
used to calculate the values of the parameters $p,q,r,s$ without
actually performing the transformation to standard form.

Though only $u_1,\ldots,u_5$ and $v_1,\ldots,v_5$ occur in
Equations~\eqref{s1s2} and \eqref{s3s4}, since the numbering is arbitrary all
six product vectors must be taken into consideration when calculating
the invariants. Different permutations of the six product vectors will
in general give different values for the invariants.

The standard form in Equation~\eqref{sformpqrs}, or equivalently the
invariants $s_i$ defined in Equations~\eqref{s1s2} and \eqref{s3s4},
can be used to check whether $\rho$ and $\rho^P$ are
$\textrm{SL}\otimes\textrm{SL}$-equivalent. We
must find the six product vectors in $\textrm{Img}\,\rho$ and
$\textrm{Img}\,\rho^P$, in some order.  Then we either transform these
to the standard form in Equation~\eqref{sformpqrs}, or calculate
directly the invariants $s_i$ for both subspaces.  For the comparison
we should try all the $6!=720$ permutations of the six product vectors
in one of the two subspaces.  If the invariants so calculated are
identical for a given permutation, then $\textrm{Img}\,\rho$ and
$\textrm{Img}\,\rho^P$ can be transformed by a unique
$\textrm{SL}\otimes \textrm{SL}$-transformation to the standard form
in Equation~\eqref{sformpqrs} with the same values of $p,q,r,s$.  The
transformations of both spaces to a common standard form then define a
unique transformation $V=V_A\otimes V_B$ from $\textrm{Img}\,\rho$ to
$\textrm{Img}\,\rho^P$.  It is then easy to check whether
$\rho^P=aV\rho_1V^{\dagger}$ for some $a>0$.

Note that different permutations may give identical sets of invariants,
but different $V$ that transform between the two spaces.  We may have
to try all of these transformations in order to find one that
transforms $\rho$ into $\rho^P$.

Note also that the partial transpose of $\rho$ with respect to
subsystem $A$ is $(\rho^P)^*$.  If $\rho$ is
$\textrm{SL}\otimes\textrm{SL}$-symmetric under this partial
transposition, then it means that the invariants of
$\textrm{Img}\,\rho^P$ will be the complex conjugates of the
invariants of $\textrm{Img}\,\rho$.  In general, this type of symmetry
is just as likely to occur.

\subsubsection*{Separable rank $(5,5)$ states}

In a generic five dimensional subspace of $\mathbb{C}^9$ containing
six normalized product vectors $w_i=u_i\otimes v_i$, we may construct
a five dimensional set of separable states as convex combinations
\begin{equation}
\rho=\sum_{i=1}^6c_i\,w_iw_i^{\dagger}
\end{equation}
with $c_i\geq 0$ and $\sum_ic_i=1$.  Hence all the separable states in
the subspace are contained in a {\emph{simplex}} with the six pure
product states as vertices.  The partial transpose of $\rho$ is
\begin{equation}
\rho^P=\sum_{i=1}^6c_i\,\widetilde{w}_i\widetilde{w}_i^{\dagger}
\end{equation}
where $\widetilde{w}_i=u_i\otimes v_i^*$ is the partial conjugate of
$w_i$.  The six partially conjugated product vectors will be linearly
independent in the generic case, hence the separable states in the
interior of the simplex will have rank $(5,6)$.

On the boundary where one coefficient $c_i$ vanishes, $\rho$ will be a
rank $(5,5)$ PPT state.  In this case five of the product vectors in
$\textrm{Img}\,\rho$ and in $\textrm{Img}\,\rho^P$ are partial
conjugates of each other, whereas the sixth product vectors in the two
spaces are related in a more complicated way, unless all the vectors
$v_i$ are real.

\subsubsection*{The surface of generic rank $(5,5)$ PPT states}

It is known that a generic four dimensional subspace is not the range
of any entangled rank $(4,4)$ PPT state~\cite{Leinaas10A}.  It appears
however in our numerical investigations that every generic five
dimensional subspace is the common range of extremal and hence
entangled rank $(5,5)$ PPT states that together form an eight
dimensional surface~\cite{Hansen12}.  It is an interesting numerical
observation that the dimension of the surface of rank $(5,5)$ PPT
states in a generic five dimensional subspace is higher than the
dimension of the simplex of separable states.

In order to compute this surface numerically we consider the
perturbation
\begin{equation}
	\label{perturbation}
	\rho'=\rho+\epsilon A
\end{equation}
with $\textrm{Tr}A=0$ and $A$ satisfying Equation~\eqref{IPQ}.

We find eight linearly independent solutions for $A$ in addition to
the trivial solution $A=\rho$, meaning that the surface has dimension
eight.

Similarly, if we want to find the dimension of the total set of rank
$(5,5)$ PPT states we find that Equation~\eqref{QQA} has 48 linearly
independent nontrivial solutions.
The dimensions 8 and 48 are consistent with the fact that
the set of five dimensional subspaces has dimension 40, see~\cite{Hansen12}.
   
\subsection{Nongeneric five dimensional subspaces}
\label{Nongeneric55170216}

By definition, for a generic set of vectors in $\mathbb{C}^3$ any
subset of three vectors will be linearly independent.  For a generic
rank $(5,5)$ PPT state $\rho$, the range ${\textrm{Img}}\,\rho$ contains six
product vectors $w_i = u_i\otimes v_i$.  A nonzero vector
$x\in\mathbb{C}^3$ can at most be orthogonal to two vectors $u_i$, and
a nonzero $y$ can at most be orthogonal to two $v_i$, hence
$z=x\otimes y$ can at most be orthogonal to four $w_i$.  Since
${\textrm{Ker}}\,\rho=({\textrm{Img}}\,\rho)^{\perp}$ it is clear that
in the generic case it is not possible to have a product vector in
$\textrm{Ker}\,\rho$, as described in Section~\ref{prodvecker}.  Thus
generic states must have $n_{\textrm{ker}}=0$.

In order to construct pairs of orthogonal subspaces
$\mathcal{U}\subset\mathbb{C}^9$ and $\mathcal{V}=\mathcal{U}^{\perp}$
with $|\mathcal{U}|=5$ and $|\mathcal{V}|=4$, such that $\mathcal{V}$
contains one or more product vectors, we must alter the generic linear
dependencies of the $u_i$ and $v_i$ vectors.  Instead of the generic
condition that any \emph{three} vectors $u_i$, and any three $v_i$,
must be linearly independent and span $\mathbb{C}^3$, we introduce the
condition that any \emph{four} vectors must span $\mathbb{C}^3$.  Then
it is possible to have one or more product vectors
$z_j=x_j\otimes y_j$ with each $x_j$ orthogonal to $u_a,u_b,u_c$ and
$y_j$ orthogonal to $v_d,v_e,v_f$ where $a,b,\ldots,f$ is some
permutation of $1,2,\ldots,6$.

In the above sense it is the \emph{subspaces} that are interesting,
and to a lesser degree the states themselves.  A general
characterization of two orthogonal subspaces $\mathcal{U}$ and
$\mathcal{V}$ with regard to the number of product vectors they
contain, is then $\{n_u;n_v\}$.

Since $z_j=x_j\otimes y_j\in\textrm{Ker}\,\rho$ if and only if
$\widetilde{z}_j=x_j\otimes y^*_j\in\textrm{Ker}\,\rho^P$, it is clear
that the kernels of $\rho$ and $\rho^P\!$ are related when they
contain product vectors.  In particular, if $y_j$ is real then
$z_j=\widetilde{z}_j\in\textrm{Ker}\,\rho^P\!$.  As long as
$n_{\textrm{ker}}\leq 4$, which is always the case in the examples we
have constructed here, we can always choose a standard form where all
the vectors $x_j$ and $y_j$ are real.


\section{A summary of numerical results}
\label{NumericalResults}

We summarize here the main results of our numerical investigations.
For ease of reference we number the cases from I to VII. 

\subsection{Generic states: Case I}

The generic rank $(5,5)$ PPT state is an extremal and hence entangled
$\{6,6;0\}$ state.  By definition, a generic state is found in a
completely random search for rank $(5,5)$ PPT states.  The number of
generic rank $(5,5)$ PPT entangled states we have generated are in the
thousands.  None of them are
$\textrm{SL}\otimes\textrm{SL}$-symmetric, and since they are extremal
they are also edge states. The partial transpose $\rho^P$ of a generic
rank $(5,5)$ PPT state $\rho$ is again a generic rank $(5,5)$ PPT state.

Unfortunately, we have made little progress in understanding these
states.  We have no general understanding of the relation between the
five dimensional subspaces $\mathrm{Img}\,\rho$ and
$\mathrm{Img}\,\rho^P$, which both contain six product vectors.  Thus
we have found no basis for a more detailed classification, in terms of
canonical forms allowing us to use analytical methods in their
construction.

\subsection{$\textrm{SL}\otimes\textrm{SL}$-symmetric states: Case Ia}

The complete set of PPT states of rank $(5,5)$ has dimension 48, and
the set of
$\mathrm{SL(3,\mathbb{C})}\otimes\mathrm{SL(3,\mathbb{C})}$-transformations
has dimension $16+16=32$, so the set of equivalence classes of rank
$(5,5)$ states with respect to
$\textrm{SL}\otimes\textrm{SL}$-transformations will have dimension
16, see~\cite{Hansen12}.  Hence in a completely random search for PPT
states of rank $(5,5)$ we will never find two states belonging to the
same equivalence class.  We also find that such a state $\rho$, which
we call generic, does never belong to the same
$\textrm{SL}\otimes\textrm{SL}$-equivalence class as its partial
transpose $\rho^P$.

This shows that $\textrm{SL}\otimes\textrm{SL}$-symmetric states can
be found numerically only by conducting restricted searches.  We will
describe now two different search methods.  For some reason that we
can only guess, the second method produces states of a more special
kind than the first method.

\subsubsection*{Method I}

An easy general procedure for constructing numerically PPT states of
specified rank that are genuinely
$\textrm{SL}\otimes\textrm{SL}$-symmetric, is to construct positive
matrices that are symmetric under partial transposition, and then
subject them to random
$\textrm{SL}\otimes\textrm{SL}$-transformations.

\subsubsection*{Method II}

Method I leaves open the question whether there exist PPT states that
are $\textrm{SL}\otimes\textrm{SL}$-symmetric but not genuinely
$\textrm{SL}\otimes\textrm{SL}$-symmetric.  In order to answer this
question, we produced in a random search more than a hundred
$\textrm{SL}\otimes\textrm{SL}$-symmetric PPT states of rank $(5,5)$,
together with their associated
$\textrm{SL}\otimes\textrm{SL}$-transformations.  We find that 50 of
these are $\{6,6;0\}$ states like the generic rank $(5,5)$ states, and
we will now describe these.  Our search was rather special in the
sense that we looked for transformations of the
form~\eqref{SL-symmetric} with $a=1$, \textit{i.e.}  transformations
that are trace preserving.  This choice allowed for the possibility
that the states found might be genuinely
$\textrm{SL}\otimes\textrm{SL}$-symmetric.

Out of the 50 states, about half are in fact genuinely
$\textrm{SL}\otimes\textrm{SL}$-symmetric, as defined by
Equation~\eqref{eq:genuineSLxSL}.  To test for genuine
$\textrm{SL}\otimes\textrm{SL}$-symmetry of a state $\rho$ we compare
the $\textrm{SL}\otimes \textrm{SL}$ transformation $V$
 from $\rho$ to
$\rho^P$, found during the search, to the special form given in
Equation~\eqref{genuineSL}.  If it is of the wanted form, we construct
the transformation $U=I\otimes U_B$ from~\eqref{eq:IsqrtVB}.  If it is
not of the wanted form, we have to produce more
$\textrm{SL}\otimes \textrm{SL}$ transformations, until we either find
one of the wanted form, or decide that probably none exists.

We consider two transformation matrices to be identical if they differ
only by a phase factor.  For the
$\textrm{SL}\otimes\textrm{SL}$-symmetric states that are not
genuinely $\textrm{SL}\otimes\textrm{SL}$-symmetric, we find only one
$\textrm{SL}\otimes \textrm{SL}$ transformation from $\rho$ to
$\rho^P$.  By definition, this is never of the form given in
Equation~\eqref{genuineSL}.  For the states of genuine
$\textrm{SL}\otimes\textrm{SL}$-symmetry found by method~II, the
number of $\textrm{SL}\otimes \textrm{SL}$ transformations we find is
always three, and out of these there is only one that has the form
given in Equation~\eqref{genuineSL}.

A special class of states proposed by Chru\'{s}ci\'{n}ski and
Kossakowski have what they call the circulant
form~\cite{Chruscinski07}
\begin{equation}
\label{ChruKossa}
\hat{\rho} =
\left(
\begin{array}{ccc|ccc|ccc}
	a_{11} & \cdot & \cdot & \cdot & a_{12} & \cdot & \cdot & \cdot & a_{13}\\
	\cdot & b_{11} & \cdot & \cdot & \cdot & b_{12} & b_{13} & \cdot & \cdot\\
	\cdot & \cdot & c_{11} & c_{12} & \cdot & \cdot & \cdot & c_{13} & \cdot\\
	\hline
	\cdot & \cdot & c_{21} & c_{22} & \cdot & \cdot & \cdot & c_{23} & \cdot\\
	a_{21} & \cdot & \cdot & \cdot & a_{22} & \cdot & \cdot & \cdot & a_{23}\\
	\cdot & b_{21} & \cdot & \cdot & \cdot & b_{22} & b_{23} & \cdot & \cdot\\
	\hline
	\cdot & b_{23} & \cdot & \cdot & \cdot & b_{32} & b_{33} & \cdot & \cdot\\
	\cdot & \cdot & c_{31} & c_{32} & \cdot & \cdot & \cdot & c_{33} & \cdot\\
	a_{31} & \cdot & \cdot & \cdot & a_{32} & \cdot & \cdot & \cdot & a_{33}\\	
\end{array}
\right).
\end{equation}
As long as the $3\times 3$ matrices $A=[a_{ij}]$, $B=[b_{ij}]$, and
$C=[c_{ij}]$ are positive, then so is $\hat{\rho}$, since the
eigenvalues of $\hat{\rho}$ are the eigenvalues of the submatrices
$A$, $B$, and $C$.  There are similar additional constraints on the
elements $a_{ij},b_{ij}$ and $c_{ij}$ in order for $\hat{\rho}$ to be
a PPT state.

Interestingly, all the genuinely
$\textrm{SL}\otimes\textrm{SL}$-symmetric PPT states of rank $(5,5)$
which we have produced by method~II, can be transformed by product
transformations to the form~\eqref{ChruKossa}.  This property is all
the more surprising, since none of the states produced by method~I
have the same property.  We will now describe how to find the
transformations.

As already mentioned, for such a state $\rho$ we always find three
different product transformations $V_1$, $V_2$, and $V_3$ such that
\begin{equation}
  V_1\rho V_1^{\dagger}=
  V_2\rho V_2^{\dagger}=
  V_3\rho V_3^{\dagger}=
  \rho^P\;.
\end{equation}
Exactly one of them is of the form~\eqref{genuineSL}.  There are six
transformations, not necessarily different, of the form
\begin{equation}
S=S_A\otimes S_B={V_j}^{-1}V_k\;,\qquad
1\leq j,k\leq 3\;,\quad
j\neq k\;,
\end{equation}
transforming from $\textrm{Img}\,\rho$ back to $\textrm{Img}\,\rho$.
It does not matter which of the six transformations we choose.  We
find that with suitable normalization the equation $S^3=I$ is
satisfied, and we have the eigenvalue decompositions
\begin{equation}
	S_A=\sum_{i=1}^3 \lambda_i\,g_ig_i^{\dagger}\;,
	\qquad
	\qquad
	S_B=\sum_{i=1}^3 \mu_i\,h_ih_i^{\dagger}\;,
\end{equation}
where the two sets of three eigenvalues $\{\lambda_i\}$ and
$\{\mu_i\}$ are both $\{1,\omega,\omega^2\}$ in arbitrary order, with
$\omega=\textrm{e}^{2\pi\textrm{i}/3}\,$.  The eigenvectors $g_i$ are
orthonormal, like the eigenvectors $h_i$.  The transformation
$T=T_A\otimes T_B$ with
\begin{equation}
	T_A^{-1}=T_A^{\dagger}={[g_1,g_2,g_3]}\;,
	\qquad
	\qquad
	T_B^{-1}=T_B^{\dagger}={[h_1,h_2,h_3]}
\end{equation}
is such that either $T\rho T^{\dagger}=\hat{\rho}$ or
$T\rho T^{\dagger}=\hat{\rho}^P$, depending on the permutation of the
eigenvectors of $S_A$ and $S_B$.

The fact that the genuinely $\textrm{SL}\otimes\textrm{SL}$-symmetric
PPT states of rank $(5,5)$ produced by method II are of a more special
type than those produced by method I, in that they can be transformed
to the circulant form~\eqref{ChruKossa}, needs an explanation.  We
guess that it is due to the existence of three product transformations
from $\rho$ to $\rho^P$, rather than one.  Presumably it is easier to
find a transformation when three solutions exist.  Hence, we introduce
a bias when we search simultaneously for a state $\rho$ of rank five
and a transformation $V=V_A\otimes V_B$ such that
$V\rho V^{\dagger}=\rho^P$.

\subsection{Nongeneric states: Cases II to VII}

Our searches for $\textrm{SL}\otimes\textrm{SL}$-symmetric states have
also produced a small number of states that have the nongeneric
property, in addition to being
$\textrm{SL}\otimes\textrm{SL}$-symmetric, that they have product
vectors in their kernels.  In particular, the very special subspace of
type $\{2;1\}$ discussed in Section~\ref{sec:Averyspecialsubspace} is
identified from four $\textrm{SL}\otimes\textrm{SL}$-symmetric states
of type $\{2,2;1\}$ found in these random searches.

In Section~\ref{Nongeneric55}, we discuss standard forms of states
that are generically not $\textrm{SL}\otimes\textrm{SL}$-symmetric,
but have the properties that $n_{\mathrm{img}}=n_u=6$ and
$n_{\mathrm{ker}}=n_v>0$.  We have generated numerically and studied
nongeneric states having from one to four product vectors in the
kernel, as discussed in Section~\ref{Nongeneric55170216}.  We define
standard forms for nongeneric orthogonal subspaces $\mathcal{U}$ and
$\mathcal{V}$ of dimensions five and four, respectively, with various
numbers $\{n_u;n_v\}$ of product vectors.  We take the product vectors
in $\mathcal{V}$ to have the standard form given in
Equation~\eqref{eq:kerz}, using only the first $n_v$ vectors.  We have
then produced a large number of PPT states $\rho$ of rank $(5,5)$ with
$\mathrm{Img}\,\rho=\mathcal{U}$ and $\mathrm{Ker}\,\rho=\mathcal{V}$.
We have studied the following nongeneric cases.

Case II, $n_{\mathrm{ker}}=1$.  Then $\textrm{Ker}\,\rho$ contains
$z_1$ from Equation~\eqref{eq:kerz}, and $\textrm{Img}\,\rho$ is
defined by Equation~\eqref{eq:61space}.  The rank $(5,5)$ PPT states
found are all extremal and of type $\{6,6;1\}$.

Case III, $n_{\mathrm{ker}}=1$.  This is the special $\{2;1\}$
subspace presented in Section~\ref{sec:Averyspecialsubspace},
producing both extremal and nonextremal entangled rank $(5,5)$ PPT
states.  The extremal states are of type $\{2,6;1\}$.  Some of the
nonextremal states are of type $\{2,6;1\}$, others are symmetric under
partial transposition and therefore of type $\{2,2;1\}$.

Case IV, $n_{\mathrm{ker}}=2$.  Then $\textrm{Ker}\,\rho$ in the
standard form contains $z_1,z_2$, and $\textrm{Img}\,\rho$ is defined
by Equation~\eqref{eq:62uv}.  The rank $(5,5)$ PPT states found are
all extremal and of type $\{6,6;2\}$.

Case V, $n_{\mathrm{ker}}=3$, where $\textrm{Ker}\,\rho$ contains
$z_1,z_2,z_3$, and $\textrm{Img}\,\rho$ is defined by
Equation~\eqref{663uv}.  The rank $(5,5)$ PPT states found are all
nonextremal and of type $\{6,6;3\}$.

Case VI, $n_{\mathrm{ker}}=4$.  This is the case discussed in
Section~\ref{sec:rank44revisited}, with $\textrm{Img}\,\rho$ defined
by Equation~\eqref{eq:64standard}.  It gives a new way of constructing
the extremal rank $(4,4)$ PPT states.

Case VII, $n_{\mathrm{ker}}=2$.  $\textrm{Ker}\,\rho$ contains
$z_1,z_2$ from Equation~\eqref{eq:kerz}, whereas $\textrm{Img}\,\rho$
is defined by Equation~\eqref{inf2uv} and contains an infinite number
of product vectors.  The resulting $\{\infty,\infty;2\}$ states
include both extremal and nonextremal states, and also some
interesting rank $(4,5)$ PPT states. We know that the latter do not
exist in generic subspaces \cite{Hansen12}.

\subsubsection*{Table}

Using the projection operators and extremality tests outlined earlier,
we can calculate the dimensions of the surfaces defined by
Equations~\eqref{extremality}--\eqref{IPQ} for the various types of
states. We use the abbreviation
${\mathbf{P}}+{\mathbf{\widetilde{P}}}$ for the operation that
preserves the range of both $\rho$ and $\rho^P$,
${\mathbf{Q}}+{\mathbf{\widetilde{Q}}}$ for the preservation of the
rank of both $\rho$ and $\rho^P$, and finally
${\mathbf{P}}+{\mathbf{\widetilde{Q}}}$ and
${\mathbf{Q}}+{\mathbf{\widetilde{P}}}$ for the two other
projections. The dimension given by $A=\rho$ has been subtracted Note
that for many of the states the dimension defined by
${\mathbf{P}}+{\mathbf{\widetilde{P}}}$ is zero, proving that the
state is extremal and therefore entangled.

Presented in Table~\ref{table:SLsym} are data for the various random
rank $(5,5)$ PPT states we have produced.  They were produced in large
numbers with the exception of the $\{2,6;1\}$ extremal states, where a
total of 15 states were produced.  As we would expect, none of the
states found were $\textrm{SL}\otimes\textrm{SL}$-symmetric, except in
case Ia where the search was restricted to
$\textrm{SL}\otimes\textrm{SL}$-symmetric states.

In addition to the dimensions defined above, we also indicate whether
the states are edge states and whether they satisfy the range
criterion, as defined in Section~\ref{subsubsecrangecrit}.
\begin{table}[H]
\begin{center}
\begin{tabular}{|c|c|c|c|c|c|c|c|}
\hline
\raisebox{0ex}[3ex][1ex]{}
Case & $\{n_{\textrm{img}},\widetilde{n}_{\textrm{img}};n_{\textrm{ker}}\}$ &
${\mathbf{Q}}+{\mathbf{\widetilde{Q}}}$ &
${\mathbf{P}}+{\mathbf{\widetilde{Q}}}$ &
${\mathbf{Q}}+{\mathbf{\widetilde{P}}}$ &
${\mathbf{P}}+{\mathbf{\widetilde{P}}}$ & \textrm{Edge} & \textrm{Range}\\
\hline
I, Ia & $\{6,6;0\}$ & 48 & 8 & 8 & 0 & \textrm{Yes} & \textrm{No}\\
II & $\{6,6;1\}$ & 49 & 9 & 9 & 0 & \textrm{Yes} & \textrm{No}\\
III & $\{2,6;1\}$ & 49 & 9 & 9 & 0 & \textrm{Yes} & \textrm{No}\\
IV & $\{6,6;2\}$ & 50 & 10 & 10 & 0 & \textrm{Yes} & \textrm{No}\\
\hline
V & $\{6,6;3\}$ & 51 & 11 & 11 & 3 & \textrm{No} & \textrm{No}\\
VI & $\{6,6;4\}$ & 52 & 12 & 12 & 5 & \textrm{No} & \textrm{Yes}\\
\hline
VII & $\{\infty,\infty;2\}$ & 50 & 12 & 12 & 0 & \textrm{Yes} & \textrm{No}\\ 
 & $\{\infty,\infty;2\}$ & 50 & 11 & 11 & 6 & \textrm{No} & \textrm{No}\\
\hline
VIIa & $\{\infty,\infty;2\}$ & 50 & 15 & 15 & 9 & \textrm{No} & \textrm{No}\\
\hline
\end{tabular} 
\end{center}
\caption{\label{table:SLsym}
  Data for the random rank $(5,5)$ PPT states we have produced based on
  the standard forms presented in Section \ref{Nongeneric55} and
  the special $\{2;1\}$ subspace in Section \ref{sec:Averyspecialsubspace}. 
  Case VIIa are PPT states of rank $(4,5)$.  For the
  dimensions of the various surfaces the trivial solution $A=\rho$
  has been subtracted, so the states with the entry zero for
  ${\mathbf{P}}+{\mathbf{\widetilde{P}}}$ are extremal.} 
\end{table}

   
\section{Nongeneric rank $(5,5)$ PPT states}
\label{Nongeneric55}

We present here several standard forms for nongeneric orthogonal
subspaces $\mathcal{U}=\textrm{Img}\,\rho$ and
$\mathcal{V}=\textrm{Ker}\,\rho$ of dimensions five and four.  The set
of parameters $a_i,b_i,c_i,d_i,e_i,f_i$ is usually assumed to be
chosen in a generic, or completely random manner. One may however also
investigate certain nongeneric choices for these coefficients.

These standard forms can be used to construct nongeneric rank $(5,5)$ PPT
states with a range defined by the given standard form, and with
$n_{\textrm{ker}}>0$.  Some of these constructions give states that
generically are extremal in $\mathcal{P}$ and therefore entangled,
while some return generically only nonextremal states.

\subsubsection*{Case V: $n_{\textrm{\rm ker}}=3$}

Given three product vectors in the kernel of $\rho$ we can make a
product transformation so that we get for these $z_i=x_i\otimes y_i$
for $i=1,2,3$ from Equation~\eqref{eq:kerz}.  We choose the
orthogonality relations
\begin{align}
\label{ref:ortho663}
x_1\perp u_i\qquad i & = 2,3,6\;,\qquad\qquad &
y_3\perp v_i\qquad i & = 1,4,5\;,
\nonumber\\
x_2\perp u_i\qquad i & = 1,3,5\;,\qquad\qquad &
y_2\perp v_i\qquad i & = 2,4,6\;,
\nonumber\\
x_3\perp u_i\qquad i & = 1,2,4\;,\qquad\qquad &
y_1\perp v_i\qquad i & = 3,5,6\;.
\end{align}
A standard form for the six product vectors $w_i = u_i\otimes v_i$ in
the range of $\rho$ is then
\begin{equation}
\label{663uv}
\settowidth{\mycolwd}{$\,\,\,a_1\,\,\,$}
u=
\left(
\begin{array}{*{6}{@{}C{\mycolwd}@{}}}
	1 & 0 & 0 & a_4 & 1 & 0\\
	0 & 1 & 0 & 1 & 0 & b_6\\
	0 & 0 & 1 & 0 & c_5 & 1
\end{array}
\right),
\qquad
v=
\left(
\begin{array}{*{6}{@{}C{\mycolwd}@{}}}
	0 & d_2 & 1 & 0 & 0 & 1\\
	1 & 0 & e_3 & 0 & 1 & 0\\
	f_1 & 1 & 0 & 1 & 0 & 0
\end{array}
\right),
\end{equation}
\begin{equation}
w=
\left(
\begin{array}{cccccc}
	\label{63space}
	0 & 0 & 0 & 0 & 0 & 0\\
	1 & 0 & 0 & 0 & 1 & 0\\
	f_1 & 0 & 0 & a_4 & 0 & 0\\

	0 & d_2 & 0 & 0 & 0 & b_6\\
	0 & 0 & 0 & 0 & 0 & 0\\
	0 & 1 & 0 & 1 & 0 & 0\\

	0 & 0 & 1 & 0 & 0 & 1\\
	0 & 0 & e_3 & 0 & c_5 & 0\\
	0 & 0 & 0 & 0 & 0 & 0
\end{array}
\right).
\end{equation} 
Let $A$ be the $6\times 6$ matrix that results from removing the rows
of zeros, \textit{i.e.} the rows one, five, and nine from $w$, then
\begin{equation}
	\label{eq:detdefabc}
	\det A=d_2e_3f_1-a_4b_6c_5\;.
\end{equation}
Since we want the six product vectors $w_i$ to span a five dimensional
space we require that $\det A=0$.  This equation determines the sixth
parameter uniquely if we choose random values for five of the
parameters $a_4,b_6,c_5,d_2,e_3,f_1$.

We impose the further restrictions that all the six parameters
$a_4,\ldots,f_1$ should be nonzero, and that
\begin{equation}
\label{663uvdet}
\settowidth{\mycolwd}{$\,\,\,a_1\,\,\,$}
\left|
\begin{array}{*{6}{@{}C{\mycolwd}@{}}}
a_4 & 1 & 0\\
1 & 0 & b_6\\
0 & c_5 & 1
\end{array}
\right|=-1-a_4b_6c_5\neq 0\;,
\qquad
\left|
\begin{array}{*{6}{@{}C{\mycolwd}@{}}}
	0 & d_2 & 1\\
	1 & 0 & e_3\\
	f_1 & 1 & 0
\end{array}
\right|=1+d_2e_3f_1\neq 0\;.
\end{equation}
Then the three vectors $u_2,u_3,u_6$ are linearly dependent, and so
are $u_1,u_3,u_5$ and $u_1,u_2,u_4$, but these are the only sets of
three vectors $u_i$ that are linearly dependent.  In the same way, the
only sets of three linearly dependent vectors $v_i$ are $v_1,v_4,v_5$,
then $v_2,v_4,v_6$, and $v_3,v_5,v_6$.  These linear dependencies make
it possible for the three product vectors $z_1,z_2,z_3$, and only
these, to be orthogonal to all the six product vectors $w_i$.  One may
check that with these choices of parameters there are no more product
vectors that are linear combinations of the six vectors $w_i$.

It should be noted that with our standard form for the three product
vectors $z_i=x_i\otimes y_i$ we still have a further freedom of doing
diagonal product transformations.  In this way we may actually reduce
the number of parameters in our standard form from six to one, setting
for example $a_4=c_5=d_2=f_1=1$.  We must then set $b_6=e_3\neq 0,-1$
in order to satisfy all the conditions.  Setting $b_6=e_3=1/p$ we then
get
\begin{equation}
\label{663uvI}
\settowidth{\mycolwd}{$\,\,\,a_1\,\,\,$}
u=
\left(
\begin{array}{*{6}{@{}C{\mycolwd}@{}}}
	1 & 0 & 0 & 1 & 1 & 0\\
	0 & 1 & 0 & 1 & 0 & 1\\
	0 & 0 & 1 & 0 & 1 & p
\end{array}
\right),
\qquad
v=
\left(
\begin{array}{*{6}{@{}C{\mycolwd}@{}}}
	0 & 1 & p & 0 & 0 & 1\\
	1 & 0 & 1 & 0 & 1 & 0\\
	1 & 1 & 0 & 1 & 0 & 0
\end{array}
\right).
\end{equation}

A PPT state $\rho$ with $\textrm{Img}\,\rho$ spanned by the product
vectors $w_1,\ldots,w_6$ and with $z_1,z_2,z_3\in\textrm{Ker}\,\rho$
must also have $z_1,z_2,z_3\in\textrm{Ker}\,\rho^P$, since
$y_1,y_2,y_3$ are real.  For given $w$ and $z$ these restrictions on
$\rho$ reduce our search for such states to a seven dimensional
subspace of $H_9$.  Six of the seven dimensions are spanned by the
pure product states $\rho_i=N_iw_iw_i^{\,\dag}$, with normalization
factors $N_i$, and there is only one dimension left where there is
room for entangled PPT states.  With $u$ and $v$ defined as in
Equation~\eqref{663uvI} the seventh direction is given by the matrix
\begin{equation}
\sigma=\left(
\begin{array}{rrrrrrrrr}
0 & 0    & 0     & 0      & \,\,\,0 & 0     & 0                   & 0           & \,\,\,0\\
0 & 0    & -\rmi & 0      & 0       & 0     & 0                   & \rmi        & 0\\
0 & \rmi & 0     & 0      & 0       & -\rmi & 0                   & 0           & 0\\
0 & 0    & 0     & 0      & 0       & \rmi  & -\rmi p^*\!\!\!\!\! & 0           & 0\\
0 & 0    & 0     & 0      & 0       & 0     & 0                   & 0           & 0\\
0 & 0    & \rmi  & -\rmi  & 0       & 0     & 0                   & 0           & 0\\
0 & 0    & 0     & \rmi p & 0       & 0     & 0                   & -\rmi p\!\! & 0\\
0 &-\rmi & 0     & 0      & 0       & 0     & \rmi p^*\!\!\!      & 0           & 0\\
0 & 0    & 0     & 0      & 0       & 0     & 0                   & 0           & 0
\end{array}
\right).
\end{equation}

In random numerical searches we have produced different sets of six
product vectors and hundreds of rank $(5,5)$ PPT states in the
corresponding seven dimensional subspace of $H_9$.  All the states we
find are entangled but not extremal.  The only extremal PPT states in
this subspace are the six pure product states $\rho_i$ and a four
dimensional surface of rank $(4,4)$ states.  The states $\rho_i$
define a five dimensional simplex of separable states, which contains
two special equilateral triangles with corners $\rho_1,\rho_2,\rho_3$
and $\rho_4,\rho_5,\rho_6$.

\begin{figure}[H]
\begin{center}
\includegraphics[width=16cm]{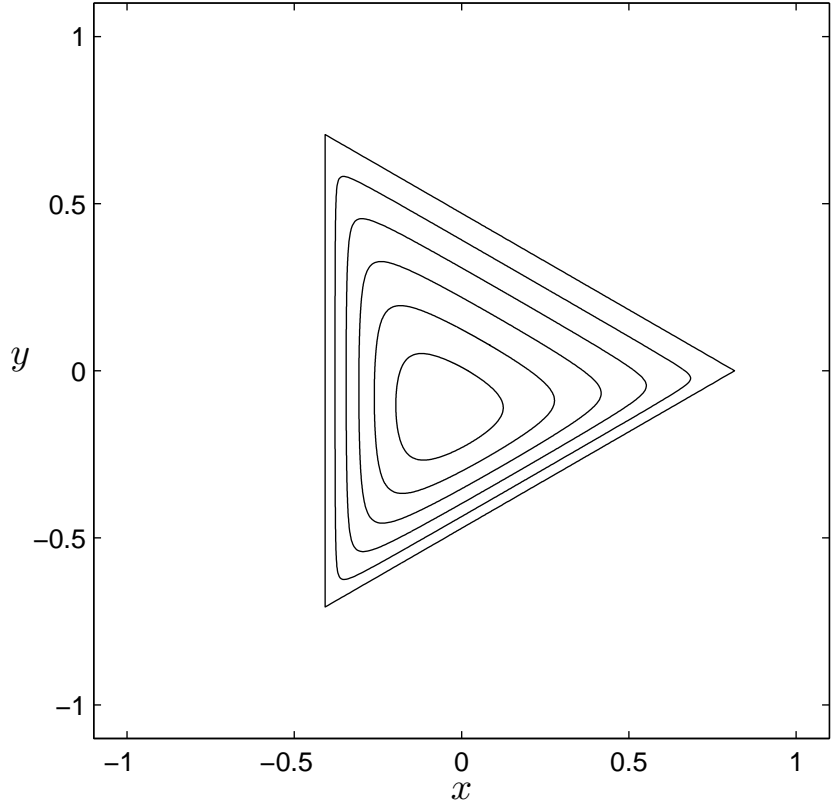}
\caption{\label{fig:contour} A three dimensional set, the interior of
  which consists of nonextremal entangled rank $(5,5)$ PPT states with
  $n_{\textrm{\rm ker}}=3$.  The set is bounded by an equilateral
  triangle at $z=0$ defined by the three pure product states
  $\rho_4,\rho_5,\rho_6$, and a two dimensional surface with $z>0$ of
  extremal rank $(4,4)$ PPT states.  The curved lines are equidistant
  contours at intervals of $\Delta z=0.07$.  The coordinates $x,y,z$
  are dimensionless.}
\end{center}
\end{figure}

A convex combination of an arbitrary rank $(4,4)$ extremal PPT state and an
arbitrary separable state from one of the two special triangles is
always a rank $(5,5)$ entangled PPT state, and every rank $(5,5)$ PPT
state of the above form can be constructed in this way.  Thus, it lies
inside a three dimensional set of rank $(5,5)$ PPT states bounded by an
equilateral triangle, either $\rho_1,\rho_2,\rho_3$ or
$\rho_4,\rho_5,\rho_6$, and a two dimensional surface of extremal rank
$(4,4)$ PPT states.  An example of this geometry is shown in
Figure~\ref{fig:contour}.  It implies that when $\rho$ is a rank
$(5,5)$ PPT entangled state there are always three product vectors
$w_i=u_i\otimes v_i\in\textrm{Img}\,\rho$ such that
$\widetilde{w}_i=u_i\otimes v_i^*\in\textrm{Img}\,\rho^P$.  The two
possible sets of three are $i=1,2,3$ or $i=4,5,6$.  Since there are
less than five such product vectors, the range criterion is not
fulfilled.

A separable state mixed from $p$ product states $\rho_i$ has rank
$(p,p)$ if $p<6$.  It has rank $(5,6)$ and lies in the interior of the
simplex if $p=6$.  Rank $(5,6)$ entangled states arise in many ways,
for example as convex combinations of one rank $(4,4)$ extremal state
and pure product states from both special triangles, for example
$\rho_1$ and $\rho_4$.  Ranks higher than six are of course
impossible, since $\textrm{Ker}\,\rho^P$ contains the three product
vectors $z_1,z_2,z_3$.  Another consequence of the last fact is that
$\textrm{Img}\,\rho^P$ contains six product vectors of the form in
Equation~\eqref{63space}, usually with other values for the
coefficients $a_4,b_6,c_5,d_2,e_3,f_1$.
 
We conclude that all of the rank $(5,5)$ states produced in these
searches are $\{6,6;3\}$ entangled nonextremal states that are neither
edge states nor satisfy the range criterion.

\subsubsection*{Case IV: $n_{\textrm{\rm ker}}=2$}

Given two product vectors in the kernel of
$\rho$ we can always make a product transformation so that we get for these
$z_i=x_i\otimes y_i$ for $i=1,2$ from Equation~\eqref{eq:kerz}.

We have the freedom of doing further product transformations with
\begin{equation}
(V_A^{\,\dag})^{-1}x_i=\alpha_i x_i\;,\qquad
(V_B^{\,\dag})^{-1}y_i=\beta_i y_i\;,\qquad
\alpha_i,\beta_i\in\mathbb{C}\;,\quad
i=1,2\;.
\end{equation}
In order to have six product vectors $w_i = u_i\otimes v_i$ orthogonal
to $z_1,z_2$, one possibility is to impose the following orthogonality
conditions,
\begin{align}
\label{eq:62overlap}
x_1\perp u_i\qquad i & = 1,2,3\;,\qquad\qquad &
y_1\perp v_i\qquad i & = 4,5,6\;,\nonumber\\
x_2\perp u_i\qquad i & = 3,4,5\;,\qquad\qquad &
y_2\perp v_i\qquad i & = 1,2,6\;.
\end{align}
These conditions are satisfied when the six product vectors
$w_i = u_i\otimes v_i$ in the range of $\rho$ have the standard form
\begin{equation}
\label{eq:62uv}
\settowidth{\mycolwd}{$\,\,\,b_1\,\,\,$}
u=
\left(
\begin{array}{*{6}{@{}C{\mycolwd}@{}}}
	0 & 0 & 0 & a_4 & a_5 & a_6\\
	b_1 & b_2 & 0 & 0 & 0 & b_6\\
	1 & 1 & 1 & 1 & 1 & 1
\end{array}
\right),
\qquad
v=
\left(
\begin{array}{*{6}{@{}C{\mycolwd}@{}}}
	d_1 & d_2 & d_3 & 0 & 0 & 0\\
	0 & 0 & e_3 & e_4 & e_5 & 0\\
	1 & 1 & 1 & 1 & 1 & 1
\end{array}
\right),
\end{equation}
with $a_i,b_i,d_i,e_i\in\mathbb{C}$.  We then get
\begin{equation}
\label{eq:62space}
\settowidth{\mycolwd}{$\,\,\,b_1d_1\,\,\,$}
w=
\left(
\begin{array}{*{6}{@{}C{\mycolwd}@{}}}
	0 & 0 & 0 & 0 & 0 & 0\\
	0 & 0 & 0 & a_4e_4 & a_5e_5 & 0\\
	0 & 0 & 0 & a_4 & a_5 & a_6\\

	b_1d_1 & b_2d_2 & 0 & 0 & 0 & 0\\
	0 & 0 & 0 & 0 & 0 & 0\\
	b_1 & b_2 & 0 & 0 & 0 & b_6\\

	d_1 & d_2 & d_3 & 0 & 0 & 0\\
	0 & 0 & e_3 & e_4 & e_5 & 0\\
	1 & 1 & 1 & 1 & 1 & 1
\end{array}
\right).
\end{equation}
We may generate random values of the coefficients $a_i,b_i,d_i,e_i$
defining any five of the six product vectors $w_i$.  Then we always
find a sixth product vector, unique up to normalization, which is a
linear combination of the five.  The complete set of six product
vectors then has rank five and automatically takes the form given
in Equation~\eqref{eq:62space}.  It is straightforward to check that $z_1,z_2$
are the only product vectors orthogonal to all the $w_i$.

We generate rank $(5,5)$ PPT states $\rho$ with
$w_1,\ldots,w_6\in\textrm{Img}\,\rho$.  With the additional
restriction that the two fixed product vectors $z_1$ and $z_2$ should
be in $\textrm{Ker}\,\rho^P$, the search can be restricted to an 11
dimensional subspace of the 81 dimensional space $H_9$ of Hermitian
matrices.

With different sets of product vectors generated in this way we have
produced numerically, by random searches, hundreds of rank $(5,5)$ PPT
states.  They are all extremal.  Hence they are also edge states, that
is, when $u_i\otimes v_i\in\textrm{Img}\,\rho$ we never have
$u_i\otimes v_i^{\,\ast}\in\textrm{Img}\,\rho^P$. The partial
transpose $\rho^P$ of such a state $\rho$ always has the same
characteristics, \textit{i.e.}\ there are six product vectors in
$\textrm{Img}\,\rho^P$ of the form given in Equation~\eqref{eq:62space}, but with
different values of the coefficients $a_i,b_i,d_i,e_i$.

In summary, all the states produced in this way are $\{6,6;2\}$
extremal states.

\subsubsection*{Case VII: $n_{\textrm{\rm ker}}=2$ with symmetric orthogonality relations}

Observe that for the above standard form given in Equation~\eqref{eq:62uv}, the
orthogonality relations in Equation~\eqref{eq:62overlap} overlap. This has been
done intentionally, and the reason will become clear when we now try
to impose the more symmetric orthogonality relations
\begin{align}
x_1\perp u_i\qquad i & = 1,2,3\;,\qquad\qquad &
y_1\perp v_i\qquad i & = 4,5,6\;,
\nonumber\\
x_2\perp u_i\qquad i & = 4,5,6\;,\qquad\qquad &
y_2\perp v_i\qquad i & = 1,2,3\;.
\end{align}
This would give six product vectors $w_i = u_i\otimes v_i$ in
$\textrm{Img}\,\rho$ on a standard form
\begin{equation}
\label{inf2uv}
u=
\left(
\begin{array}{cccccc}
	0 & 0 & 0 & a_4 & a_5 & a_6\\
	b_1 & b_2 & b_3 & 0 & 0 & 0\\
	1 & 1 & 1 & 1 & 1 & 1
\end{array}
\right),
\qquad
v=
\left(
\begin{array}{cccccc}
	d_1 & d_2 & d_3 & 0 & 0 & 0\\
	0 & 0 & 0 & e_4 & e_5 & e_6\\
	1 & 1 & 1 & 1 & 1 & 1
\end{array}
\right),
\end{equation}
where $a_i,b_i,d_i,e_i\in\mathbb{C}$.  As compared
to Equation~\eqref{eq:61space} it means that we set
$b_4=b_5=b_6=e_1=e_2=e_3=0$.  We then get
\begin{equation}
w=
\left(
\begin{array}{cccccc}
	\label{inf2space}
	0 & 0 & 0 & 0 & 0 & 0\\
	0 & 0 & 0 & a_4e_4 & a_5e_5 & a_6e_6\\
	0 & 0 & 0 & a_4 & a_5 & a_6\\

	b_1d_1 & b_2d_2 & b_3d_3 & 0 & 0 & 0\\
	0 & 0 & 0 & 0 & 0 & 0\\
	b_1 & b_2 & b_3 & 0 & 0 & 0\\

	d_1 & d_2 & d_3 & 0 & 0 & 0\\
	0 & 0 & 0 & e_4 & e_5 & e_6\\
	1 & 1 & 1 & 1 & 1 & 1
\end{array}
\right).
\end{equation} 
Omitting one of the six product vectors gives ${\textrm{rank}}\,w=5$.
However, omitting for example $w_6$, we can find no sixth product
vector of the form $w_6$ in the five dimensional subspace.  On the
other hand, there will be infinitely many product vectors that are
linear combinations of $w_1,w_2,w_3$.  In fact, $u_3$ and $v_3$ must
be linear combinations
\begin{equation}
u_3=\alpha_1u_1+\alpha_2u_2\;,\qquad
v_3=\beta_1v_1+\beta_2v_2\;,
\end{equation}
and then for any $\gamma_1,\gamma_2\in\mathbb{C}$ the product vector
$w'=u'\otimes v'$ with
\begin{equation}
u'=\gamma_1\alpha_1u_1+\gamma_2\alpha_2u_2\;,\qquad
v'=\gamma_1\beta_1v_1+\gamma_2\beta_2v_2
\end{equation}
will be a linear combination of $w_1,w_2,w_3$.  All these product
vectors lie in a three dimensional subspace of a
$\mathbb{C}^2\otimes\mathbb{C}^2$ subspace.

Our search for rank $(5,5)$ PPT states $\rho$ with
$w_1,\ldots,w_5\in\textrm{Img}\,\rho$ and
$z_1,z_2\in\textrm{Ker}\,\rho^P$ can now be restricted to a 13
dimensional subspace of $H_9$.  Using five such randomly generated
product vectors $w_1,\ldots,w_5$ to create subspaces of dimension
five, we have produced numerically by random searches hundreds of
rank $(5,5)$ PPT states.  Most of the states we find are separable, but we
also find a very small number of extremal rank $(5,5)$ PPT states.

We find that 11 of the 13 dimensions in $H_9$ represent unnormalized
separable states, and the entangled PPT states account for the last
two of the 13 dimensions.  This can be understood as follows.  The
product vectors $w_1,w_2,w_3$ span a three dimensional subspace
containing infinitely many product vectors.  These product vectors
generate a set of unnormalized separable states of dimension nine, the
same as the complete set of unnormalized density matrices on the three
dimensional subspace.  In addition we get two more dimensions of
separable states from the product vectors $w_4$ and $w_5$.

Note that a separable state mixed from more than three product vectors
in the $\mathbb{C}^2\otimes\mathbb{C}^2$ subspace will have rank
$(3,4)$.  Hence, if we mix in also one or both of $w_4,w_5$ we will
get separable states of rank $(4,5)$ or rank $(5,6)$, respectively.

The partial transposes of the states we construct numerically have the
same characteristics, thus all the states are of type
$\{\infty,\infty;2\}$.  It is noteworthy that it is possible for an
extremal rank $(5,5)$ PPT state to have infinitely many product
vectors in its range.

\subsubsection*{Case II: $n_{\textrm{\rm ker}}=1$}

Given a product vector in the kernel of $\rho$ we
may perform a product transformation as in Equation~\eqref{SL-transformation1}
so that the vector after normalization take the form $z_1=x_1\otimes y_1$ from Equation~\eqref{eq:kerz}.

This transformation is not unique, and we have the freedom of doing
further transformations with
\begin{equation}
(V_A^{\,\dag})^{-1}x_1=\alpha x_1\;,\qquad
(V_B^{\,\dag})^{-1}y_1=\beta y_1\;,\qquad
\alpha,\beta\in\mathbb{C}\;.
\end{equation}
We assume that the six product vectors $w_i = u_i\otimes v_i$ in the
range of $\rho$ have the standard form
\begin{equation}
\label{eq:61space}
u=
\left(
\begin{array}{cccccc}
	0 & 0 & 0 & a_4 & a_5 & a_6\\
	b_1 & b_2 & b_3 & b_4 & b_5 & b_6\\
	1 & 1 & 1 & 1 & 1 & 1
\end{array}
\right),
\qquad
v=
\left(
\begin{array}{cccccc}
	d_1 & d_2 & d_3 & 0 & 0 & 0\\
	e_1 & e_2 & e_3 & e_4 & e_5 & e_6\\
	1 & 1 & 1 & 1 & 1 & 1
\end{array}
\right)
\end{equation}
where $a_i,b_i,d_i,e_i\in\mathbb{C}$. They are then orthogonal to
$z_1 = x_1\otimes y_1$ because
\begin{align}
x_1\perp u_i\qquad i & = 1,2,3\;,\qquad\qquad &
y_1\perp v_i\qquad i & = 4,5,6\;.
\end{align}
The normalization in Equation~\eqref{eq:61space} presupposes that the third
component of every vector is nonzero.  There is no loss of generality
in this assumption, because of the freedom we have to do product
transformations.  This freedom could actually be used to reduce the
number of parameters in the standard form given in Equation~\eqref{eq:61space}.

In our numerical searches we generate, for example, random values for
the coefficients $a_i,b_i,d_i,e_i$ for $i=1,\ldots,5$.  This is then a
generic case in which $w_1,\ldots,w_5$ span a five dimensional
subspace, and there will be a unique solution for $a_6,b_6,e_6$ such
that $w_6$ lies in this subspace.  There are no other product vectors
in the subspace, in the generic case.

Furthermore, there can be no product vector $z'=x'\otimes y'$
orthogonal to all $w_i$, apart from $z'=z_1$.  In fact, $x'$ can at
most be orthogonal to three vectors $u_i$, because any four of the
vectors $u_i$ span $\mathbb{C}^3$.  Similarly, $y'$ can at most be
orthogonal to three vectors $v_i$.  The only set of three linearly
dependent vectors $u_i$ is $u_1,u_2,u_3$, and the only set of three
linearly dependent vectors $v_i$ is $v_4,v_5,v_6$.  The only
possibility is therefore that $x'\perp u_i$ with $i=1,2,3$ and
$y'\perp v_i$ with $i=4,5,6$, implying that $z'=z_1$.

We generate rank $(5,5)$ PPT states with this subspace as range.  In
general, there is a 25 dimensional real vector space of Hermitian
matrices operating on a fixed subspace of dimension five.  However,
when we are searching for a PPT state $\rho$ with $\rho z_1=0$,
we have the additional
restriction that the partially conjugated product vector
$\widetilde{z}_1=x_1\otimes y_1^{\ast}=x_1\otimes y_1=z_1$ must be in
$\textrm{Ker}\,\rho^P$.  This reduces the 25 dimensions to 17, because
the equation $\rho^P\!z_1=0$ gives the four extra complex equations
$\rho_{jk}=0$ with $j=2,3$ and $k=4,7$.  It is straightforward to find
numerically rank $(5,5)$ PPT states in this 17 dimensional space of
Hermitian matrices.

In random searches we have produced hundreds of such states for many
different sets of product vectors $w_i$.  All states produced in this
manner are extremal.  Hence they are edge states, that is, when
$u_i\otimes v_i\in\textrm{Img}\,\rho$ we never have
$u_i\otimes v_i^{\,\ast}\in\textrm{Img}\,\rho^P$.  We find that the
partial transpose $\rho^P$ always has the same characteristics as
$\rho$, \textit{i.e.}\ there are six product vectors in
$\textrm{Img}\,\rho^P$ of the form in Equation~\eqref{eq:61space}, but
with different values for the coefficients $a_i,b_i,d_i,e_i$.

So in our classification all these states are $\{6,6;1\}$ extremal
states.
 

\section{Case III: An exceptional subspace of type $\{2;1\}$}
\label{sec:Averyspecialsubspace}

A generic five dimensional subspace of
$\mathbb{C}^3\otimes\mathbb{C}^3$ contains exactly six product
vectors.  Obviously, a nongeneric five dimensional subspace may well
contain infinitely many product vectors, if it contains a whole
product space $\mathbb{C}^1\otimes\mathbb{C}^2$,
$\mathbb{C}^2\otimes\mathbb{C}^1$, or
$\mathbb{C}^2\otimes\mathbb{C}^2$.  On the other hand, it may also
contain less than six product vectors, if the set of equations
in~\eqref{productvectors} have degenerate solutions~\cite{Chen11}.
The maximum dimension of an entangled subspace, {\em i.e.}\ a subspace
that does not contain any product vectors, is known from
\cite{Parthasarathy04}, and for the $3 \times 3$ system the limiting
dimension is four. So any subspace of ${\mathbb{C}^9}$ of dimension
five or higher \emph{must} contain at least one product vector. We
describe here a very special $\{2;1\}$ subspace of ${\mathbb{C}^9}$.

In our random searches for rank $(5,5)$ PPT states with the special
symmetry property that the state $\rho$ and its partial transpose
$\rho^P$ are $\textrm{SL}\otimes\textrm{SL}$-equivalent, we have found
four such states with $n_{\textrm{img}}=2$.  A further special
property of these states is that $n_{\textrm{ker}}=1$, \textit{i.e.}
there is exactly one product vector in ${\textrm{Ker}}\,\rho$.  The
states are not extremal, since three of them lie on a line between a
pure product state and a rank $(4,4)$ extremal PPT state, whereas one
lies inside a two dimensional region bounded by a closed curve of rank
$(4,4)$ extremal PPT states.

Further numerical random searches for rank $(5,5)$ PPT states with the
same ranges as the nonextremal rank $(5,5)$ PPT states described
above, but not restricted to $\textrm{SL}\otimes\textrm{SL}$-symmetric
states, reveal extremal rank $(5,5)$ PPT states of type $\{2,6;1\}$.
It is clear that a subspace with only two product vectors supports
just one line of separable states.

We find that the pure product state and the curve of rank $(4,4)$
states are parts of the same geometry, which can be transformed to a
standard form as shown in Figure~\ref{fig:5}.  It is an empirical fact
that a rank $(4,4)$ state $\rho_{44}$ involved in such a construction of a
rank $(5,5)$ state $\rho$ can always be transformed to a standard form
where the product vectors in ${\textrm{Ker}}\,\rho_{44}$ are related
to regular icosahedra~\cite{Sollid11}. We will now describe this
standard form in more detail.

\subsection{Rank $(5,5)$ PPT states in a region bounded by rank $(4,4)$ PPT states}

We describe here analytically a set of states found by transformation
to standard form of one particular rank $(5,5)$ nonextremal PPT state
found numerically in a random search.  This particular state lies
inside a curve of rank $(4,4)$ states, but the three other states we
have found belong to the same geometry.  It is remarkable that
analytically defined states possessing such very special properties
turn up in random searches.  Our limited imagination would not have
enabled us to deduce their existence.

After transformation to standard form, the rank $(5,5)$ state lies inside a
circle bounded by extremal rank $(4,4)$ PPT states.  The interior of
the circle consists entirely of rank $(5,5)$ PPT states, each of which
has exactly one product vector in its kernel, this vector is common to
all the rank $(4,4)$ and rank $(5,5)$ states.  Each of the rank $(4,4)$
states has five additional product vectors in its kernel, these are
different for the different states.  The rank $(4,4)$ states have no
product vectors in their ranges, whereas all the rank $(5,5)$ states have
one common range containing exactly two product vectors.  All the
rank $(4,4)$ and rank $(5,5)$ states are symmetric under partial transposition.

\subsubsection*{Product vectors from a regular icosahedron}

In the standard form we define, all the product vectors in the kernels
of all the rank $(4,4)$ states are defined from regular icosahedra, as
follows.  We define $c_k=\cos(2k\pi/5)$, $s_k=\sin(2k\pi/5)$, thus
\begin{align}
    c_1 & = c_4 = \frac{\sqrt{5}-1}{4}\;, &
    s_1 & = -s_4 = \frac{\sqrt{10+2\sqrt{5}}}{4}\;,\nonumber\\
    c_2 & = c_3 = -\frac{\sqrt{5}+1}{4}\;, &
    s_2 & = -s_3 = \frac{\sqrt{10-2\sqrt{5}}}{4}\;.
\end{align}
Note that $\phi=2c_1=0.61803\ldots$ is the golden mean, defined by the
equation $\phi^2=1-\phi$.

We define product vectors $z_k=x_k\otimes y_k$ for $k=1,2,\ldots,6$
with
\begin{equation}
\settowidth{\mycolwd}{$\,\,2c_1\,\,$}
\label{xy}
x=\frac{1}{\sqrt{5}}
\left(
\begin{array}{*{6}{@{}C{\mycolwd}@{}}}
	2 & 2c_1 & 2c_2 & 2c_3 & 2c_4 & 0\\
	0 & 2s_1 & 2s_2 & 2s_3 & 2s_4 & 0\\
	1 & 1 & 1 & 1 & 1 & \sqrt{5}
\end{array}
\right),\;\;\;\;
y=\frac{1}{\sqrt{5}}
\left(
\begin{array}{*{6}{@{}C{\mycolwd}@{}}}
	2c_2 & 2 & 2c_3 & 2c_1 & 2c_4 & 0\\
	2s_2 & 0 & 2s_3 & 2s_1 & 2s_4 & 0\\
	1 & 1 & 1 & 1 & 1 & \sqrt{5}
\end{array}
\right).
\end{equation}
The 12 vectors $\pm x_k$ are real and are all the corners of a regular
icosahedron.  The vectors $y_k$ as defined here are the same vectors
in a different order.  We define $x$ and $y$ in such a way that
$x_5=y_5$ and $x_6=y_6$.  The product vector $z_6$ is going to play a
special role, it will be the one common product vector in the kernels
of any two of the rank $(4,4)$ states.

\subsubsection*{A circle of extremal rank $(4,4)$ PPT states}

The six product vectors $z_k$ are linearly dependent and define a five
dimensional subspace of $\mathbb{C}^9$.  The orthogonal projection on
this subspace may be written as
\begin{equation}
Q=\frac{5}{6}\sum_{k=1}^6 z_k{z_k}^{\!\dag}\;.
\end{equation}
Since $Q$ is a projection, with $Q^2=Q$, and is symmetric under
partial transposition, $Q^P=Q$, we conclude that the matrix
\begin{equation}
\tau_{44}=\frac{1}{4}\,({I}-Q)
\end{equation}
is a rank $(4,4)$ PPT state, entangled and extremal.  This method for
constructing such states is known as a UPB construction, since any
five of the six vectors $z_k$ form an unextendible product basis for
the subspace, so called because there is no product vector orthogonal
to all of them~\cite{Bennett99,DiVincenzo03}.

Note that $Q$ and $\tau_{44}$ are symmetric under partial
transposition with respect to subsystem $B$, in our notation $Q^P=Q$
and $\tau_{44}^{\,P}=\tau_{44}$, because the vectors $y_k$ defined
in Equation~\eqref{xy} are real.  They are also symmetric under partial
transposition with respect to subsystem $A$, in our notation
$Q^{PT}=Q$ and $\tau_{44}^{\,PT}=\tau_{44}$, because the vectors $x_k$
are real.

We now define product vectors $w_k=u_k\otimes v_k$ for $k=1,2$ with
\begin{equation}
\label{u1v1u2v2}
u_1=\frac{1}{\sqrt{2}}\left(
\begin{array}{c}
 1\\-\mathrm{i}\\0
\end{array}\right)\;,
\qquad
u_2=\left(
\begin{array}{c}
 0\\0\\1
\end{array}\right)\;,
\qquad
v_1=v_2^{\,\ast}=u_1^{\,\ast}\;.
\end{equation} 
Note that $u_2=x_6=y_6$ and
\begin{equation}
u_1=\frac{1}{\sqrt{10}}\sum_{j=1}^5 \omega^{j-1}\,x_j
\qquad
\mbox{with}
\qquad
\omega=\textrm{e}^{-2\pi\textrm{i}/5}=c_1-\textrm{i}s_1\;.
\end{equation}
We define
\begin{equation}
W=w_1{w_2}^{\!\dag}
=(u_1{u_2}^{\!\dag})\otimes(v_1{v_2}^{\!\dag})\;,
\end{equation}
and then the two Hermitian matrices
\begin{equation}
A=W+W^{\dag}\;,\qquad
B=\textrm{i}\,(W-W^{\dag})\;.
\end{equation}
Since $v_1=v_2^{\,\ast}$, $W$ is symmetric under partial
transposition,
\begin{equation}
W=(u_1{u_2}^{\!\dag})\otimes(v_1v_1^{\,T})
=(u_1{u_2}^{\!\dag})\otimes(v_1v_1^{\,T})^T
=W^P\;,
\end{equation}
and so are $A$ and $B$.  Define now
\begin{equation}
\label{21range}
\rho=\tau_{44}+\alpha A+\beta B\;.
\end{equation}
with real parameters $\alpha$ and $\beta$.  The eigenvalues of $\rho$
are four times 0, three times 1/4, and
\begin{equation}
\lambda_{\pm}
=\frac{1}{8}\pm
\frac{1}{24}\,\sqrt{1+
 \big(24\alpha-2\sqrt{2}\,c_1\big)^2
+\big(24\beta-2\sqrt{2}\,s_1\big)^2}\;.
\end{equation}
Since $\rho^P=\rho$, we get a rank $(4,4)$ PPT state with
$\lambda_-=0$ when
\begin{equation}
\label{eq:circle}
 \big(24\alpha-2\sqrt{2}\,c_1\big)^2
+\big(24\beta-2\sqrt{2}\,s_1\big)^2=8\;.
\end{equation}
In particular, $\alpha=\beta=0$ gives the state $\tau_{44}$ that we
started from.  Equation~\eqref{eq:circle} defines a circle of rank $(4,4)$
states $\rho_{44}$, which are in fact extremal PPT states.  All the states inside
the circle are rank $(5,5)$ nonextremal PPT states.  The centre of
the circle we call $\rho_0$, it has $\alpha=\sqrt{2}\,c_1/12$ and
$\beta=\sqrt{2}\,s_1/12$.

All the states defined by Equation~\eqref{21range} are symmetric under partial
transposition with respect to subsystem $B$, we have that
\begin{equation}
\rho^P
=\tau_{44}^{\,P}+\alpha A^P+\beta B^P
=\tau_{44}+\alpha A+\beta B
=\rho\;.
\end{equation}
However, because $A$ and $B$ are complex, $\rho$ is complex when
$\alpha\neq 0$ or $\beta\neq 0$.  Then it is not symmetric under
partial transposition with respect to subsystem $A$, we have then that
\begin{equation}
\rho^{PT}=\rho^T=\rho^{\ast}\neq\rho\;.
\end{equation}
 
All the rank $(4,4)$ states $\rho_{44}$ on the circle are projections.
The rank $(4,4)$ state at an angle $\gamma$ around the circle from the
state $\tau_{44}$ is $\rho_{44}=U\tau_{44}U^{\dagger}$ where
$U=U_A\otimes U_B$ is a unitary product transformation,
\begin{equation}
\label{eq:UAUB}
U_A=\left(
\begin{array}{ccc}
\cos(\gamma/10) & -\sin(\gamma/10) & 0\\
\sin(\gamma/10) &  \cos(\gamma/10) & 0\\
0 & 0 & {\rm e}^{-{\rm i}\gamma/2}
\end{array}
\right),
\quad
U_B=\left(
\begin{array}{ccc}
\cos(\gamma/5) & \sin(\gamma/5) & \phantom{--}0\\
-\sin(\gamma/5) &  \cos(\gamma/5) & \phantom{--}0\\
0 & 0 & \phantom{--}1
\end{array}
\right).
\end{equation}
The kernel of $U\tau_{44}U^{\dagger}$ is defined by the transformed
icosahedron vectors $(U_A^{\dag})^{-1}x=U_Ax$ and
$(U_B^{\dag})^{-1}y=U_By$.  This transformation leaves $y_6$ invariant
and $x_6$ invariant up to a phase factor.  A rotation by $\gamma=2\pi$
is a cyclic permutation of the first five icosahedron vectors,
\begin{equation}
x_1\mapsto -x_4\mapsto x_2\mapsto -x_5\mapsto x_3\mapsto -x_1\;,\qquad
y_1\mapsto y_4\mapsto y_2\mapsto y_5\mapsto y_3\mapsto y_1\;.
\end{equation}
Note that the vectors $u_1$, $u_2$, $v_1$, and $v_2$ are eigenvectors of
the transformation matrices $U_A$ and $U_B$, for example,
\begin{equation}
U_Au_1=\textrm{e}^{{\rm i}\;\!\gamma/10}\,u_1\;.
\end{equation}
Hence the pure states $\rho_k=w_kw_k^{\dag}$ are invariant under the
transformation $\rho\mapsto U\rho U^{\dag}$.

\begin{figure}[H]
\begin{center}
\includegraphics[width=10cm]{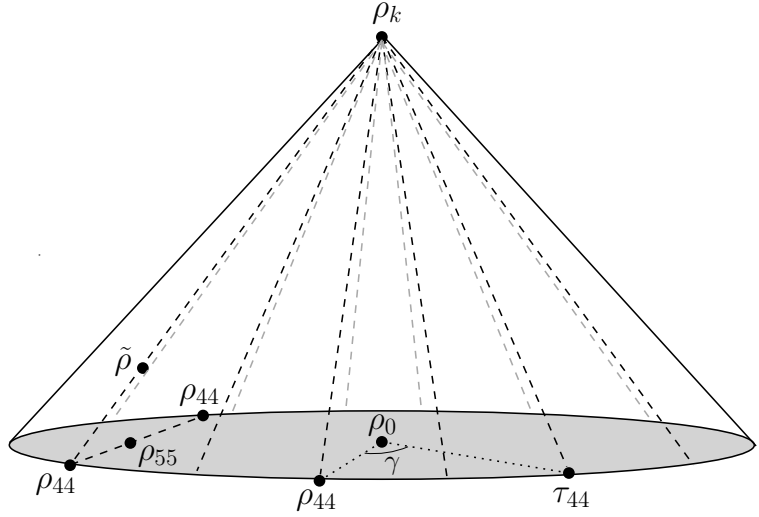}
\caption{\label{fig:5} A pure product state
  $\rho_k=w_k{w_k}^{\dagger}$ with $k=1$ or $k=2$, together with the
  circle of extremal rank $(4,4)$ PPT states, define a three-dimensional
  conical face of $\mathcal{P}$.  The state $\rho_0$ is defined as the
  centre of the circle.}
\end{center}
\end{figure}

\subsubsection*{Two three-dimensional solid cones}

The geometry of this construction is depicted in Figure \ref{fig:5}.
The four traceless matrices $A,B,\rho_1-\rho_0$, and $\rho_2-\rho_0$
define directions in $H_9$ that are mutually orthogonal.  The
directions $A$ and $B$ define the plane of the circle centred on
$\rho_0$.  Thus, the two states $\rho_k=w_k{w_k}^{\dagger}$ with
$k=1,2$, together with the circle of rank $(4,4)$ states, define two cones
that are different three-dimensional faces of $\mathcal{P}$, the set
of PPT states.  All the nonextremal states on the surface of a cone
are rank $(5,5)$ states.  A state $\rho$ in the interior of the cone may be
written as a convex combination
\begin{equation}
\rho=p\rho_{55}+(1-p)\rho_k\;,\qquad 0<p<1\;,
\end{equation}
with $\rho_{55}$ inside the circle and $k=1,2$.  This state also
has rank five, since $w_k\in\textrm{Img}\,\rho_{55}$.  But its
partial transpose $\rho^P$ will have rank six, because it is a convex
combination of $\rho_{55}^{\,P}=\rho_{55}$ and one of the two
pure product states
\begin{equation}
\rho_1^P=(u_1\otimes v_2)(u_1\otimes v_2)^{\dag}\;,\qquad
\rho_2^P=(u_2\otimes v_1)(u_2\otimes v_1)^{\dag}\;.
\end{equation}
Since $\rho_{55}^{\,PT}=\rho_{55}^T\neq\rho_{55}$, that is,
$\rho_{55}$ is not symmetric under partial transposition with
respect to system $A$, we conclude that neither is $\rho$.  The same
conclusion, that $\rho^{PT}\neq\rho$, follows because $\rho^{PT}$ must
have rank six, in fact it has the same eigenvalues as $\rho^P$.

\subsubsection*{The conical surface}

Since $u_2$ as defined in Equation~\eqref{u1v1u2v2} is real, it follows that
$\rho^{PT}=\rho$ when we define
\begin{equation}
\label{eq:partsym1}
  \rho=p\tau_{44}+(1-p)\rho_2\;,\qquad 0<p<1\;.
\end{equation}
Thus, a state $\rho$ of this form is a rank $(5,5)$ state on the
surface of the cone that is symmetric under partial transposition with
respect to system $A$.  Since $\rho^P$ and $\rho^{PT}$ have the same
eigenvalues, it follows that $\rho$ and $\rho^P$ have the same
eigenvalues, although they are neither $\textrm{SU}\otimes\textrm{SU}$-equivalent nor
$\textrm{SL}\otimes\textrm{SL}$-equivalent.

If we define
\begin{equation}
\label{eq:partsym1a}
  \widetilde{\rho}=p\rho_{44}+(1-p)\rho_2\;,
\end{equation}
using a different rank $(4,4)$ state $\rho_{44}$ in place of the special
state $\tau_{44}$, then we get a state $\widetilde{\rho}$ that has no
longer the same symmetry as $\rho$, but is equivalent to $\rho$ by a
unitary product transformation $U$ as given in
Equation~\eqref{eq:UAUB}.  In this case again $\widetilde{\rho}$,
$\widetilde{\rho}^P$, and $\widetilde{\rho}^{PT}$ all have the same
eigenvalues, in fact, they have the same eigenvalues as $\rho$,
$\rho^P$, and $\rho^{PT}$.

If we replace $\rho_2$ by $\rho_1$ in Equation~\eqref{eq:partsym1},
then the state $\rho$ we get is not
$\textrm{SL}\otimes\textrm{SL}$-equivalent to any one of $\rho^P$ or
$\rho^{PT}$.  We find empirically that all three states $\rho$,
$\rho^P$, and $\rho^{PT}$ still have the same set of eigenvalues,
although the reason is not clear.

\subsubsection*{Four similar constructions}

The construction just described is one of four different ways of
extending the four dimensional subspace $\textrm{Img}\,\tau_{44}$ to a
five dimensional subspace which is orthogonal to $z_6$ and contains
exactly two product vectors $w_1,w_2$.  Define
\begin{equation}
a=\left(
\begin{array}{c}
 0\\0\\1
\end{array}\right)\;,
\qquad
b=\frac{1}{\sqrt{2}}\left(
\begin{array}{c}
 1\\\mathrm{i}\\0
\end{array}\right),
\end{equation} 
We find that the only possibilities are those given in
Table~\ref{tab:4poss}.  They give rise to four different circles of
rank (4,4) extremal PPT states.  The specific possibility given by
Equation~\eqref{u1v1u2v2} is case 2 in Table~\ref{tab:4poss}. Note
that the cases~1 and~2 are related by complex conjugation, and so
are~3 and~4.

\begin{table}[H]
\begin{center}
\begin{tabular}{|@{}l|r@{$\otimes$}l|r@{$\otimes$}l|}
\hline
\multicolumn{1}{|c|}{Case} & \multicolumn{2}{c|}{$w_1$} & \multicolumn{2}{c|}{$w_2$}\\
\hline
\,\,\,\,\,\, 1 & $b\,$ & $\,b^*$ & $a\,$ & $\,b$\\
\,\,\,\,\,\, 2 & $b^*$ & $\,b$ & $a\,$ & $\,b^*$\\
\,\,\,\,\,\, 3 & $b\,$ & $\,b$ & $b^*$ & $\,a$\\
\,\,\,\,\,\, 4 & $b^*$ & $\,b^*$ & $b\,$ & $\,a$\\
\hline
\end{tabular}
\end{center}
\caption{\label{tab:4poss}The four ways to extend the
  four dimensional subspace $\textrm{Img}\,\tau_{44}$ to a five
  dimensional subspace orthogonal to $z_6$ and containing exactly
  two product vectors $w_1,w_2$.}
\end{table}

In our numerical random searches for
$\textrm{SL}\otimes\textrm{SL}$-symmetric PPT states of rank $(5,5)$
we have found four examples of states that are
$\textrm{SL}\otimes\textrm{SL}$-equivalent to states lying on the
surface of such a cone.  In one example the state lies inside the
circle, and is of the form given in Equation~\eqref{21range}.  This is
the case discussed in detail above, corresponding to case~2 in
Table~\ref{tab:4poss}.  In the three other examples the state lies on
the conical surface, and is of the form given in
Equation~\eqref{eq:partsym1}.  These correspond to the cases~3 and~4
in Table~\ref{tab:4poss}, where $\rho_2^P=\rho_2$.


\section{Summary and outlook}
\label{sec:summaryandoutlook}

The work presented here is mainly a continuation of previous studies
of the entangled PPT states of rank five in the $3\times 3$ system,
with an emphasis on nongeneric states.

For dimension $3\times 3$ it is known that the extremal PPT states of
lowest rank, apart from the pure product states, are entangled states
of rank four.  It is by now well established that all these rank four
states are equivalent by
$\textrm{SL}\otimes\textrm{SL}$-transformations to states constructed
by simple procedures that can be described analytically.  The first
such procedure, introduced by Bennett \emph{et al.}, was based on
unextendible product bases~\cite{Bennett99}.  A second procedure is
described in~\cite{Hansen12}.  As a byproduct of our present study of
rank $(5,5)$ states we have found a third procedure, which is
described here.  The structure of extremal rank five PPT states is
much more complex, and no exhaustive classification or set of
construction methods is known.
 
The equivalence between PPT states under
$\mathrm{SL}\otimes\mathrm{SL}$-transformations is an important
concept, and quantities invariant under such transformations are
especially useful for classification.  In particular, we classify a
state $\rho$ by the ranks $(m,n)$ of $\rho$ and $\rho^P$, and by the
number of product vectors
$\{n_{\textrm{img}},\widetilde{n}_{\textrm{ker}};n_{\textrm{img}}\}$
in the subspaces $\mathrm{Img}\,\rho$, $\mathrm{Img}\,\rho^P$, and
$\mathrm{Ker}\,\rho$, respectively.  Generic PPT states of rank
$(5,5)$ are extremal and of type $\{6,6;0\}$.

As a first attempt towards a general classification we define a
state $\rho$ to be $\mathrm{SL}\otimes\mathrm{SL}$-symmetric under
partial transposition if it is
$\mathrm{SL}\otimes\mathrm{SL}$-equivalent to its partial transpose
$\rho^P$.  We define $\rho$ to be \emph{genuinely}
$\mathrm{SL}\otimes\mathrm{SL}$-symmetric if it is
$\mathrm{SL}\otimes\mathrm{SL}$-equivalent to a state $\tau$ with
$\tau=\tau^P\!$.  We then show that genuine
$\mathrm{SL}\otimes\mathrm{SL}$-symmetry implies
$\mathrm{SL}\otimes\mathrm{SL}$-symmetry, in such a way that at least
one $\mathrm{SL}\otimes\mathrm{SL}$-transformation from $\rho$ to
$\rho^P$ must have a special diagonal block form and in addition be
trace preserving.

In random numerical searches where we searched specifically for
$\mathrm{SL}\otimes\mathrm{SL}$-symmetric PPT states of rank $(5,5)$,
along with their associated transformations, we found about 50 such
states of type $\{6,6;0\}$.  The condition of
$\mathrm{SL}\otimes\mathrm{SL}$-symmetry enforced during the search
means that these states are not generic.  Another restriction imposed
was that we looked for product transformations that were trace
preserving.  This allowed us to find states that are genuinely
$\mathrm{SL}\otimes\mathrm{SL}$-symmetric.
Out of the 50 states produced in this way, about half are genuinely
$\mathrm{SL}\otimes\mathrm{SL}$-symmetric.
All these genuinely $\mathrm{SL}\otimes\mathrm{SL}$-symmetric states
are $\mathrm{SL}\otimes\mathrm{SL}$-equivalent to a special class of
states proposed by Chru\'{s}ci\'{n}ski and
Kossakowski~\cite{Chruscinski07}.  This is a curious result which we
do not understand, especially since random states of rank five that
are constructed with the only restriction that they should be
symmetric under partial transposition, can not be transformed to this
special form.

How to construct PPT states with $n_{\mathrm{ker}}>0$ was discussed
in~\cite{Hansen12}, and we develop these matters further here.  For
the $3\times 3$ system this is essentially a study of orthogonal
complementary subspaces $\mathcal{U},\mathcal{V}\subset\mathbb{C}^9$
of dimensions five and four, with certain nongeneric properties as
described in Section~\ref{Nongeneric55170216}.  We have constructed
several standard forms for $\mathcal{U}$ and $\mathcal{V}$, with from
one to four product vectors in $\mathcal{V}$, and then produced random
PPT states of rank $(5,5)$ with $\mathrm{Img}\,\rho=\mathcal{U}$ and
$\mathrm{Ker}\,\rho=\mathcal{V}$.  A detailed summary is given in
Section~\ref{NumericalResults}.

For the case $n_{\mathrm{ker}}=4$ we find the new analytical
construction, already mentioned, of all rank four extremal PPT states,
up to $\mathrm{SL}\otimes\mathrm{SL}$-equivalence, where they appear
as boundary states on one single five dimensional face on the set of
normalized PPT states. The interior of the face consists of rank five
states, a simplex of separable states surrounded by entangled PPT
states. All these states are real matrices, symmetric under partial
transposition.

A very special subspace, of type $\{2;1\}$, is collected from a small
number of $\{2,2;1\}$ states found in our random searches for
$\mathrm{SL}\otimes\mathrm{SL}$-symmetric states.  We describe
analytically a set of states on such a subspace, in a standard form,
illustrated in Fig.~\ref{fig:5}.

Apart from this special case, we have only considered five dimensional
subspaces containing at least six product vectors.  The maximum
dimension of an entangled subspace, containing no product vectors, is
known from \cite{Parthasarathy04}.  For the $3 \times 3$ system the
limiting dimension is four, hence any subspace of dimension five or
higher must contain at least one product vector.  An analysis on how
to construct subspaces with fewer than six product vectors is given
in~\cite{Chen11}.  To fully describe extremal rank five PPT states
according to the number of product vectors in the range and kernel,
all these cases should be investigated.

Finally, we want to remind the reader that the basic problem of
understanding the generic PPT states of rank $(5,5)$ remains unsolved.
Also, we are even further from a full understanding of higher rank
extremal PPT states in $3\times 3$ dimensions, or in higher
dimensions.


\section*{Acknowledgements}

We acknowledge gratefully the research grant from 
The Norwegian University of Science and Technology (Leif Ove Hansen) and 
cooperation and help from 
B\o rge Irgens especially with development of programs, 
but also with other insights during the research.  



\begin{thebibliography}{99}

\bibitem{Bell64}
J.S. Bell,\\
\emph{On the Einstein Podolsky Rosen paradox}.\\
Physics \textbf{1}, 195 (1964).

\bibitem{Aspect82}
A. Aspect, J. Dalibard, and G. Roger,\\
\emph{Experimental test of Bell's inequalities using time-varying analyzers}.\\
Phys. Rev. Lett. \textbf{49}, 25 (1982).

\bibitem{Gharibian10}
S. Gharibian,\\
\emph{Strong NP-hardness of the quantum separability problem}.\\
Quant. Inf. and Comp. \textbf{10}, 3 (2010).

\bibitem{MPRHorodecki96}
M. Horodecki, P. Horodecki, and R.Horodecki,\\
\emph{Separability of mixed states: necessary and sufficient conditions}.\\
Phys. Lett. A \textbf{223}, 1 (1996).

\bibitem{Peres96}
A. Peres,\\
\emph{Separability criterion for density matrices}.\\
Phys. Rev. Lett \textbf{77}, 1413 (1996).

\bibitem{PHorodecki97}
P. Horodecki,\\
\emph{Separability criterion and inseparable mixed states with positive partial transposition}.\\
Phys. Lett. A \textbf{232}, 5 (1997).

\bibitem{PHorodecki00}
P. Horodecki, M. Lewenstein, G. Vidal, and I. Cirac,\\
\emph{Operational criterion and constructive checks for the separability of low-rank density matrices}.\\
Phys. Rev. A \textbf{62}, 032310 (2000).

\bibitem{Bennett99}
C.H. Bennett, D.P. DiVincenzo, T. Mor, P.W. Shor, J.A. Smolin, and B.M. Terhal,\\
\emph{Unextendible Product Bases and Bound Entanglement}.\\
Phys. Rev. Lett. \textbf{82}, 5385 (1999).

\bibitem{DiVincenzo03}
D.P. DiVincenzo, T.Mor, P.W. Shor, J.A. Smolin, and B.M. Terhal,\\
\emph{Unextendible Product Bases, Uncompletable Product Bases and Bound Entanglement}.\\
Commun. Math. Phys. \textbf{238}, 379 (2003).

\bibitem{Leinaas10A}
J.M. Leinaas, J. Myrheim, and P.\O. Sollid,\\
\emph{Low-rank extremal positive-partial-transpose states and unextendible product bases}.\\
Phys. Rev. A \textbf{81}, 062330 (2010).

\bibitem{Chen11}
L. Chen and D.Z. Djokovic,\\
\emph{Description of rank four PPT entangled states of two qutrits}.\\
J. Math. Phys. \textbf{52}, 122203 (2011).


\bibitem{Leinaas07}
J.M. Leinaas, J. Myrheim, and E. Ovrum,\\
\emph{Extreme points of the set of density matrices with positive partial transpose}.\\
Phys. Rev. A \textbf{76}, 034304 (2007). 

\bibitem{Leinaas10B}
J.M. Leinaas, J. Myrheim, and P.\O. Sollid,\\
\emph{Numerical studies of entangled PPT states in composite quantum systems}.\\
Phys. Rev. A \textbf{81}, 062329 (2010). 

\bibitem{Hartshorne}
R. Hartshorne,\\
\emph{Algebraic Geometry}.\\
Springer, New York (2006).

\bibitem{Parthasarathy04}
K.R. Parthasarathy,\\
\emph{On the maximal dimension of a completely entangled subspace for finite level quantum systems},\\
Proc. Math. Sci. \textbf{114}, 464 (2004).

\bibitem{Skowronek11}
{\L}. Skowronek,\\
\emph{Three-by-three bound entanglement with general unextendible product bases.}\\
J. Math. Phys. \textbf{52}, 122202 (2011).

\bibitem{Hansen12}
L.O. Hansen, A. Hauge, J. Myrheim, and P.\O. Sollid,\\
\emph{Low-rank positive-partial-transpose states and their relation to product vectors}.\\
Phys. Rev. A \textbf{85}, 022309 (2012).

\bibitem{Chruscinski07}
D. Chru\'{s}ci\'{n}ski and A. Kossakowski,\\
\emph{Circulant states with positive partial transpose}.\\
Phys. Rev. A \textbf{76}, 032308 (2007).

\bibitem{Sollid11}
P.{\O}. Sollid, J.M. Leinaas, and J. Myrheim,\\
\emph{Unextendible product bases and extremal density matrices with positive partial transpose}.\\
Phys. Rev. A \textbf{84}, 042325 (2011).

\end{thebibliography}
\end{document}